\documentclass[useAMS,usenatbib]{mn2e}
\usepackage{amsmath,amssymb,amsfonts} 
\usepackage{graphics}                     
\usepackage{color}                        
\usepackage{graphicx}
\usepackage{mathrsfs}  
\usepackage[authoryear]{natbib}
\usepackage{hyperref}                     
\DeclareFontFamily{U}{euc}{}
\DeclareFontShape{U}{euc}{m}{n}{<-6>eurm5<6-8>eurm7<8->eurm10}{}%
\DeclareSymbolFont{AMSc}{U}{euc}{m}{n} 
\DeclareMathSymbol{\upmu}{\mathord}{AMSc}{"16}

\title{Extension of an Exponential Light Curve GRB Pulse Model Across Energy Bands}
\author[R. J. Nemiroff]{Robert J. Nemiroff$^{1}$\footnotemark[1]\\
$^{1}$Department of Physics, Michigan Technological University, Houghton, MI 49931}
\begin{document}

\date{\date{Accepted ... Received ... ; in original form ...}}
 
\pagerange{\pageref{firstpage}--\pageref{lastpage}} \pubyear{2011}
 
\maketitle

\label{firstpage}
 
\begin{abstract}
A simple mathematical model of GRB pulses in time, suggested in \citet{Nor05}, is extended across energy.  For a class of isolated pulses, two of those parameters appear effectively independent of energy.  Specifically, statistical fits indicate that pulse amplitude $A$ and pulse width $\tau$ are energy dependent, while pulse start time and pulse shape are effectively energy independent.  These results bolster the Pulse Start and Pulse Scale conjectures of \citet{Nem00} and add a new Pulse Shape conjecture which states that a class of pulses all have the same shape.  The simple resulting pulse counts model is $P(t,E) = A(E)  \ {\rm exp} (-t/\tau(E) - \tau(E)/t)$, where $t$ is the time since the start of the pulse.  This pulse model is found to be an acceptable statistical fit to many of the fluent separable BATSE pulses listed in \citet{Nor05}.  Even without theoretical interpretation, this cross-energy extension may be immediately useful for fitting prompt emission from GRB pulses across energy channels with a minimal number of free parameters. 
\end{abstract}
 
\begin{keywords}
Gamma-Rays: Bursts - Gamma-Rays: observations
\end{keywords}

\section{Introduction}
\label{sec:introduction}

Gamma Ray Burst (GRB) prompt emission appears typically to be composed of distinct emission episodes known as ``pulses" \citep{Des81, Nor96}.  Although pulses typically overlap in time, a fraction of GRBs feature a pulse bright enough, long enough, and separate enough to be analyzed by itself.  Previously, several authors have suggested relatively simple analytic forms for GRB pulses (e.g. \citet{Nor96, Ryd02, Nor05}), and several conjectures have been given and tested involving GRB pulse coherence \citep{Nem00, Hak09}.

The \citet{Nor96} and \citet{Nor05} mathematical forms fit GRB pulse light curves to an exponential rise and exponential decay, while the \citet{Ryd02} form fits GRB pulse light curves to a power law rise and power law decay. \citet{Nem11} fit a Planckian functional form that has an exponential rise and a power-law decay. \citet{Sch96} showed that 10 GRB pulses detected early in the Burst and Transient Source Experiment's (BATSE's) mission, on board the now-defunct Compton Gamma Ray Observatory, were only marginally well fit to an exponential decay, while a power-law decay sometimes fits better.

Here a simple analytical pulse models suggested previously by \citet{Nor05} is analyzed in greater detail.  In particular, the pulse model is reformulated mathematically into three component parameters: one that scales linearly in time, one that scales linearly in intensity, and one that solely determines the pulse model shape.  These parameters easily scale with energies at which the pulse is observed.

All previously published pulse fitting schemes attempted to fit a pulse only at a single energy or in a single energy band.  Fits to the same pulse at another energy are typically started fresh, with all the free parameters again being determined from scratch. To date, no published system uses information from a pulse at one energy to fit the same pulse at another energy. Deconvolving complicated GRBs into pulses, however, can be a computationally expensive procedure \citep{Hak08}.  The problem is not just an inefficient use of computer time -- it affects fitting accuracy as well -- information gained from fitting the pulse in a bright energy channel is not being used to formally constrain the fit of a dim energy channel, although it may be used as a starting point \citep{Hak08}.

In Section 2 the mathematical \citet{Nor05} pulse model is reviewed and reformulated in terms of more easily scalable parameters.  Section 3 shows how the scalable \citet{Nor05} pulse is extendable across energy channels.  Section 4 shows fits for the energy generalized pulse model to several bright individual BATSE pulses selected from \citet{Nor05}.  Section 5 details a search for correlations between scalable parameters found in the fits, while in Section 6 some discussions and conclusions are given.

\section{The Norris Pulse Model}

\subsection{The Norris Paramatrization}

In \citet{Nor05}, the following pulse model was fit to numerous GRB pulses and discussed: 
  \begin{equation} \label{NorrisRaw}
  P(t) = A \ e^{-\tau_r/t - t/\tau_d } ,
  \end{equation}
where $P$ is the count rate of the pulse in counts per second, $t$ is time during the GRB, $A$ is the amplitude of the pulse also in counts per second, $\tau_r$ is a temporal factor that scales the rise of the pulse, and $\tau_d$ is a temporal factor that scales the decline of the pulse. Here the pulse start time $t_0$ has implicitly been set to zero.  Although the first author of this paper, J. P. Norris, has been involved in the creation and testing of other pulse models, here this pulse model will be referred to as the ``Norris pulse model". In \citet{Nor05}, pulses were fit separately to the Norris pulse model in every energy band.

As with any peaked function, the time of peak counts is found simply by finding when $dP/dt = 0$.  For the Norris pulse model this occurs when $t = t_{peak} = \sqrt{\tau_r \tau_d}$.  At this time, the maximum count rate of the pulse is 
  \begin{equation}
  P_{peak} = P(t_{peak}) = A \ e^{-2 \sqrt{\tau_r / \tau_d}}  .
  \end{equation}
The pulse shape can be uniquely determined by also finding $P$ at another time, here chosen to be at twice the time between the peak and the start of the pulse, so that  
  \begin{equation}
  P(2 t_{peak}) = A \ e^{-(5/2) \sqrt{\tau_r / \tau_d}} .
  \end{equation}

\subsection{A Different Parametrization of the Norris Pulse Model}

The same pulse shape can be rewritten in terms of other variables with the goal of giving greater insight into the cross-energy scaling potentially inherent in GRB pulses.  Specifically, Eq.(\ref{NorrisRaw}) can be rewritten as
  \begin{equation} \label{NorrisScale}
  P(t) = A e^{-\xi (t/\tau + \tau/t)}  .
  \end{equation}
Since $\tau = \sqrt{\tau_r \tau_d}$ and $\xi = \sqrt{\tau_r / \tau_d}$, it is clear that this pulse description involves just a direct variable substitution and so is mathematically identical to the previously defined Norris Pulse Model. This pulse light curve parametrization appears more easily scalable than the original parametrization, a feature that will be discussed in some detail below.

In this parametrization, the peak of the pulse occurs at $t_{peak} = \tau$ where 
  \begin{equation}
  P_{peak} = P(t_{peak}) = A e^{-2 \xi} ,
  \end{equation}  
so that 
  \begin{equation}
  \xi = -(1/2) {\rm ln} (P(t_{peak})/A).  
  \end{equation}
Note that this does not uniquely specify $\xi$ since the value of $A$ remains unknown, and the value of $t_{peak}$ is known only relative to the start time of the pulse: $t_o$.

At twice the peak time, however, 
 \begin{equation}
  P_2 = P(2 t_{peak}) = A e^{-5 \xi /2 } ,
  \end{equation}
so that $P_{peak} / P_2 = e^{-\xi / 2}$.  Since this ratio eliminates $A$, it is now possible to show that 
  \begin{equation}
  \xi = -2 \ {\rm ln} (P_{peak} / P_2) ,
  \end{equation}
which does uniquely specify $\xi$ in terms of the relative value intensity of the pulse peak to the intensity at twice the time of the peak.

\begin{figure}
\includegraphics[width=84mm]{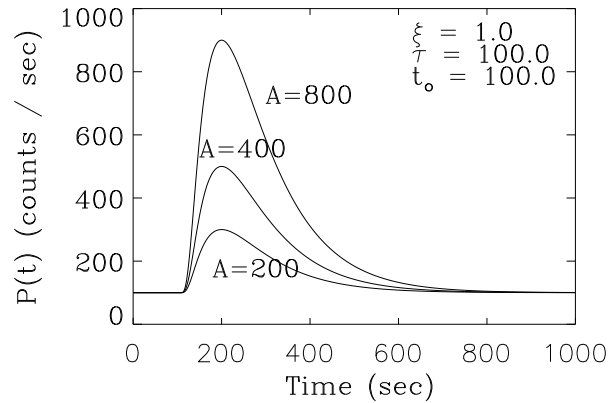}
\caption{A plot of count rate versus time for theoretical Norris pulse model for different amplitudes $A$. Here the pulses share all other defining parameters, specifically $\xi=1$, $\tau=100$, and $t_o=100$ sec. An artificial background level of 100 counts per second was input. Note how changing $A$ only scales the count rate axis (here the y-axis), while leaving all other aspects of the pulse unchanged.}   
\label{aplot}
\end{figure}

\begin{figure}
\includegraphics[width=84mm]{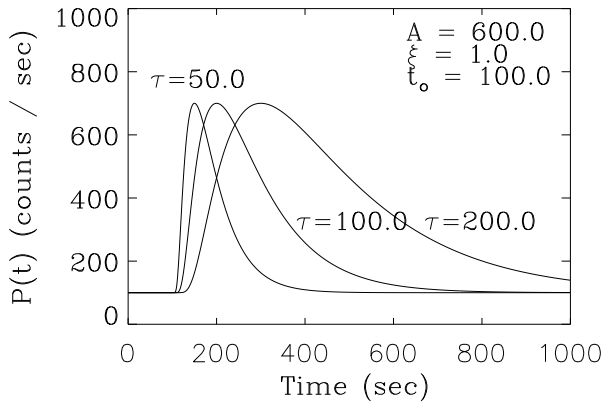}
\caption{A plot of count rate versus time for theoretical Norris pulse model for different time scaling factors $\tau$. Here the pulses share all other defining parameters, specifically $\xi=1$, $A=60$, and $t_o=100$ sec. An artificial background level of 100 counts per second was input. Note how changing $\tau$ only scales the time axis (here the x-axis), while leaving all other aspects of the pulse unchanged.}
\label{tplot}
\end{figure}

This scalable parametrization allows the deconvolution of the pulse shape into three parameters that have separable and effectively orthogonal meanings.  First, scaling $A$ will stretch the counts  (``y-axis") of the pulse only, leaving the timing (``x-axis") and the inherent shape of the pulse untouched.  This is shown graphically in Figure (\ref{aplot}).  Next, scaling $\tau$ will stretch only the time axis of the pulse, leaving the amplitude and the inherent shape of the pulse untouched.  This is shown graphically in Figure (\ref{tplot}).  Note that this time stretching occurs while holding the pulse start time $t_o$ fixed.

\begin{figure}
\includegraphics[width=84mm]{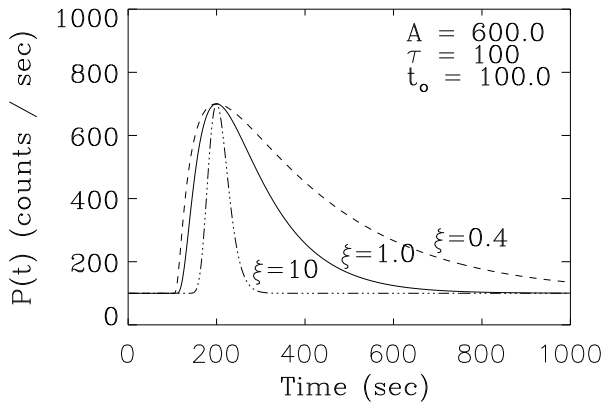}
\caption{A plot of count rate versus time for theoretical Norris pulse model for different pulse shape parameters $\xi$. Here the pulses share all other defining parameters, specifically $A=60$, $\tau=100$, and $t_o=100$ sec. An artificial background level of 100 counts per second was input. Note how changing $\xi$ fundamentally changes the shape of the pulse. Note that lower values of $\xi$ create more asymmetric pulses, while higher values of $\xi$ create more symmetric pulses.}
\label{splot}
\end{figure}

Last, in this deconvolution, the $\xi$ parameter alone defines the scale-independent shape of the pulse.  Only when $\xi$ is changed, for example, can the asymmetry of the pulse be changed.  This is shown graphically in Figure (\ref{splot}). Changing the shape parameter $\xi$, however, will also change the amplitude of the pulse.  The only remaining free parameter is $t_o$, the start time of the pulse.

Note that in this parametrization, $\tau$, the time of the peak relative to the start time of the pulse, can be used as a proxy for width parameter $w$ used in \citet{Nor05}.  As shown in Figure (4) of \citet{Nor05}, there is excellent agreement between the two.

Since $t_{peak} = \tau$, a straightforward method for estimating $\tau$ is between the measured time values of the pulse peak and $t_0$.  Although measuring the peak time is relatively straightforward, measure the start time is more complex.  However, since most pulses rise relatively rapidly compared to their decay, \citet{Hak09}, for example, found it possible to estimate $t_0$values to an accuracy below 0.4 seconds from a statistical fit of the Norris pulse form.

\section{Extension of Norris Pulse Model to Other Energies}

The scalable parametrization of the Norris pulse model can be easily generalized to other energies, so long as the pulse light curve is fit acceptably at those energies.  Generally,  Eq. (\ref{NorrisScale}) can be expanded so that 
  \begin{equation} \label{NorrisFull}
  P(t, E) = {A(E) \over e^{2 \xi(E)}}  \
  e^{ -\xi(E) \left( {(t - t_o(E)) \over  \tau(E)} + {\tau(E) \over (t - t_o(E))} \right) } ,
  \end{equation}
where it is now assumed that $A$, $\tau$, $\xi$, and $t_o$ are all functions of energy $E$.

This generalization would be particularly useful if any of the functions $A(E)$, $\xi(e)$, $\tau(E)$, or $t_o(E)$ were known since that would reduce the number of free parameters needed to fit a pulse.  Such a function might come from theory one day, but presently one can postulate a few empirical rules bounding these functions from the inspection of numerous GRB light curves.  One such empirical hypothesis is that the values of these functions are smoothly varying over $E$, so that the knowledge of a functional value at one energy is likely similar to a value at a close energy.  Past this, however, only two concrete relations have been hypothesized to relate GRB pulse light curves between energies: the Pulse Start Conjecture \citep{Nem00} and the Pulse Scale Conjecture \citep{Nem00}.

The Pulse Start Conjecture posits that every pulse has the same start time at every energy, the mathematical statement of which is that $t_o(E) = t_o$, a constant that is not a function of energy. The Pulse Start Conjecture was posed and tested in \citet{Nem00} for the most fluent pulse in BATSE trigger 2193 (GRB 930214c) for 16 MER energy channels and found to be an acceptable statistical fit for all of them.  Further, the Pulse Start Conjecture was tested in \citet{Nem00} using BATSE four-energy channel data for the most fluent pulses in four other BATSE long GRBs (GRB 930612a, GRB 940529b, GRB 941031b, and GRB 970825, corresponding to BATSE trigger numbers 2387, 3003, 3267, and 6346, respectively) and found to be statistically consistent.

However, \citet{Nor05} measured $t_o$ values for the brightest pulse in four energy channels for BATSE GRB 2193 while fitting the pulse.  The most discrepant fit was found to be start times that differed by 2.0 $\sigma$ from each other.  A statistically significant difference was recorded for BATSE trigger 2387 between energy channels 1 and 2, and channels 1 and 3, where the \citet{Nor05} $t_o$ values differed by 4.7 $\sigma$ and 8.2 $\sigma$ respectively.  However, these differences could be due to a single displacement in the channel one start time for this pulse. More generally, \citet{Nor05} measured start values for numerous bright pulses in BATSE GRBs and recorded data which include cases that appear to follow the Pulse Start Conjecture, and some cases that don't, assuming that the error in $t_o$ measured is gaussian.

Most recently, \citet{Hak09} tested 199 pulses found in 75 long GRBs and and found that the Pulse Start Conjecture to be generally plausible, with a few errant cases possibly attributable to secondary pulses or low signal data.

Given that the Pulse Start Conjecture holds, then Eq. (\ref{NorrisFull}) can be simplified to
  \begin{equation} \label{NorrisStart}
  P(t, E) = {A(E) \over e^{2 \xi(E)}} 
  e^{ -\xi(E) \left( {t \over  \tau(E)} + {\tau(E) \over t} \right) } ,
  \end{equation}
where $t$ is now considered measured relative to an energy independent $t_o$.

A separate conjecture, the Pulse Scale Conjecture, posits that GRB pulses differ between energies only by scale factors in time and flux \citep{Nem00}.  In this paradigm, it is $\xi$ that is constant and hence no longer a function of energy.  Note that the Pulse Scale Conjecture allows different pulses to have different shapes.  Given that Pulse Scale Conjecture holds, then Eq. (\ref{NorrisFull}) can be further simplified to
  \begin{equation} \label{NorrisScale2}
 P(t,E) = A(E) e^{ -\xi (t / \tau(E) + \tau(E) / t) } .
  \end{equation}
The remaining energy dependant terms are now $A(E)$ and $\tau(E)$, with $A(E)$ now usefully being the ``y-axis" flux scale factor and $\tau(E)$ being the othogonal``x-axis" temporal scale factor.

A yet further conjecture ventured here, the Pulse Shape Conjecture, posits that a group of GRB pulses all have the same shape.  In this paradigm, $\xi$ is not only constant in a pulse across energies, an entire group of pulses all are hypothesized to have the same shape: $\xi = 1$.  Given that the Pulse Shape Conjecture holds, then Eq. (\ref{NorrisFull}) can be further simplified to   
\begin{equation} \label{NorrisShape}
 P(t,E) = A(E) e^{ -(t / \tau(E) + \tau(E) / t) } .
\end{equation}
It is suspected that this functional form is not unique in that other mathematical functions can also adequately fit all of the pulse conjectures and describe the same group of GRB pulses.

In the above cross-energy pulse descriptions, the meaning of $A(E)$ is interesting as it appears to be a type of spectrum for a pulse that involves the flux from the entire pulse.  Note that $A(E)$ will be different than a spectrum taken at a specific time for of the pulse, say at the time of peak energy of one of the energy channels.  Interestingly, $A(E)$ can be fit from incomplete pulse information, for example computable when only a fraction of a single pulse is recoverable across energy bands.

Previously, $A(E)$ has only been determined for a single pulse in a single GRB: the main pulse in BATSE trigger 2193 \citep{Nem00}.  There, $A(E$) was fit to 16 BATSE MER channels and was seen to peak at an energy of about 150 keV. Note that $A(E)$ is intrinsically a type of ``photon count spectrum", as $P(t)$ is a measurement of photon counts rather than energy flux.

The meaning of $\tau(E)$ appears to be the pulse duration as a function of energy.  Note that $\tau(E)$ has (also) only been determined for single pulse in a single GRB, in fact the same pulse (BATSE trigger 2193) as with $A(E)$ \citep{Nem00}.  There, $\tau(E)$ appears approximated as a power law where $\tau(E) \propto E^{-0.7}$.  Informal observation of numerous GRB pulses indicates that $\tau(E)$ usually decreases monotonically with increasing energy.  Note that as with $A(E)$, $\tau(E)$ may also be fit from incomplete pulse information, for example computable when only a faction of a single pulse is recoverable across energy bands.

Note that the total fluence of a Norris-model pulse is directly proportional to $A(E)$ times $\tau(E)$.  Since $A(E)$ has a clear peak, pulses might be seen as envelopes effectively spanning only certain times and energies.

\section{Fitting GRB Pulses with Constraints Across Energy Channels}

In this section, an attempt is made to determine whether the Norris pulse model can be constrained by the Pulse Start, Pulse Scale, and Pulse Shape conjectures and still adequately fit isolated GRB pulses adequately across BATSE energy bands. Note that invoking these conjectures provides many additional constraints, so that many fewer free parameters will be available to fit the pulses across the BATSE energy bands. As a consequence, one might expect fits to be less good than fits without these constraints. Therefore, what is being tested is not whether such constraints provide better fits -- but whether these more tightly constrained models can create acceptable fits at all.

BATSE data was used for several reasons.  First, BATSE incorporated the largest set of GRB detectors ever flown, and so recorded relatively high count rates for GRBs. Next, BATSE, so far, was one of the longest running GRB detectors ever flown, and so over its decade of operation had time for wait for relatively bright GRBs to occur. Next, BATSE data has now been around for over two decades and so is relatively well understood.  Next, BATSE data is in the public domain and easily available over the web in simple formats, for example ASCII. Last, \citet{Nor05} created their bright isolated pulse catalog from BATSE data, which is a catalog with useful data from which this study can draw and compare.

\begin{table*}
Table 1: Temporal Fit Parameters
\begin{center}
\begin{tabular}{| c | c | c | c | c | c | c |}
\hline\hline
GRB &  BATSE Trigger & Energy Range &         $t_o$ &  $A$ & $\tau$ & $\chi^2/\nu$  \\
\hline
930214B & 2193 & 20 - 50       & -5.9 &  257  & 23.0 & 1.58 \\
930214B & 2193 & 50 - 100     & -5.9 &  724  & 20.9 & 1.55 \\
930214B & 2193 & 100 - 300   & -5.9 & 1450 & 15.1 & 1.36 \\
\hline
930217 & 2197 & 20 - 50       & -2.3 $\pm$ 0.5 & 182 $\pm$ 27 & 10.96 $\pm$ 1.6 & 1.09 \\
930217 & 2197 & 50 - 100     & -2.3 $\pm$ 0.5 & 195 $\pm$ 21 &  7.59 $\pm$ 0.35 & 1.19 \\
930217 & 2197 & 100 - 300   & -2.3 $\pm$ 0.5 & 316 $\pm$ 23 & 5.25 $\pm$ 0.25 & 1.30 \\
\hline
930612 & 2387 & 20 - 50       & -1.8 &  724  & 10.0  & 1.49 \\
930612 & 2387 & 50 - 100     & -1.8 & 1020 &  8.71 & 1.18 \\
930612 & 2387 & 100 - 300   & -1.8 & 1180 & 6.92  & 1.88 \\
\hline
931128 & 2665 & 20 - 50       & -1.2 $\pm$ 0.3 & 170 $\pm$ 25 & 3.98 $\pm$ 0.39 & 1.20 \\
931128 & 2665 & 50 - 100     & -1.2 $\pm$ 0.3 & 275 $\pm$ 20 &  2.88 $\pm$ 0.14 & 1.00 \\
931128 & 2665 & 100 - 300   & -1.2 $\pm$ 0.3 & 224 $\pm$ 16 & 1.91 $\pm$ 0.09 & 1.33 \\
\hline
941023B & 3256 & 20 - 50       & -0.8 $\pm$ 0.4 & 52.5 $\pm$ 12 & 5.50 $\pm$ 1.74 & 1.15 \\
941023B & 3256 & 50 - 100     & -0.8 $\pm$ 0.4 & 105 $\pm$ 15  &  3.63 $\pm$ 0.54 & 1.07 \\
941023B & 3256 & 100 - 300   & -0.8 $\pm$ 0.4 & 129 $\pm$ 19  & 2.40 $\pm$ 0.23 & 1.08 \\
\hline
941026 & 3257 & 20 - 50       & -2.2 &  138  & 10.96  & 1.53 \\
941026 & 3257 & 50 - 100     & -2.2 &  295 &   9.12 & 2.41 \\
941026 & 3257 & 100 - 300   & -2.2 &  417 & 6.61  & 3.09 \\
\hline
960331B & 5387 & 20 - 50       & -2.5 $\pm$ 0.6 & 91.2 $\pm$ 29   & 8.32 $\pm$ 3.70 & 0.99 \\
960331B & 5387 & 50 - 100     & -2.5 $\pm$ 0.6 & 182 $\pm$ 13    &  6.92 $\pm$ 0.67 & 1.27 \\
960331B & 5387 & 100 - 300   & -2.5 $\pm$ 0.6 & 275 $\pm$ 20     & 4.37 $\pm$ 0.64 & 0.89 \\
\hline
960409C & 5415 & 20 - 50       & -1.4 $\pm$ 0.4 & 240 $\pm$ 36 & 7.24 $\pm$ 0.35 & 1.29 \\
960409C & 5415 & 50 - 100     & -1.4 $\pm$ 0.4 & 390 $\pm$ 28  &  5.25 $\pm$ 0.50 & 1.00 \\
960409C & 5415 & 100 - 300   & -1.4 $\pm$ 0.4 & 390 $\pm$ 28  & 3.89 $\pm$ 0.48 & 0.94 \\
\hline
970330A & 6147 & 20 - 50       & -2.6 $\pm$ 0.7 & 105 $\pm$ 16 & 7.59 $\pm$ 1.96 & 0.99 \\
970330A & 6147 & 50 - 100     & -2.6 $\pm$ 0.7 & 182 $\pm$ 13  & 6.31 $\pm$ 0.61 & 1.26 \\
970330A & 6147 & 100 - 300   & -2.6 $\pm$ 0.7 & 138 $\pm$ 20  & 4.57 $\pm$ 0.68 & 1.04 \\
\hline
\hline
980213A & 6598 & 20 - 50       & -1.3 &  402  & 6.20  & 2.68 \\
980213A & 6598 & 50 - 100     & -1.3 &  430 &  5.16 & 2.64 \\
980213A & 6598 & 100 - 300   & -1.3 &  350 &  3.75  & 1.52 \\
\hline
980302 & 6625 & 20 - 50       & -2.0 &  493  & 7.44  & 3.12 \\
980302 & 6625 & 50 - 100     & -2.0 &  566 &  6.79 & 2.42 \\
980302 & 6625 & 100 - 300   & -2.0 &  350 &  5.40  & 1.12 \\
\hline
980425 & 6707 & 20 - 50       & -3.4 &  249  & 9.34  & 0.98 \\
980425 & 6707 & 50 - 100     & -3.4 &  306 &  7.44 & 1.63 \\
980425 & 6707 & 100 - 300   & -3.4 &  267 & 5.66  & 1.38 \\
\hline
980913 & 7087 & 20 - 50       & -3.6 $\pm$ 0.9 & 217 $\pm$ 49 & 8.15 $\pm$ 0.35 & 0.99 \\
980913 & 7087 & 50 - 100     & -3.6 $\pm$ 0.9 & 306 $\pm$ 22  & 7.10 $\pm$ 0.34 & 1.26 \\
980913 & 7087 & 100 - 300   & -3.6 $\pm$ 0.9 & 327 $\pm$ 48  & 6.20 $\pm$ 0.59 &  1.13 \\
\hline
981015A & 7156 & 20 - 50       & -2.6 &  249  & 8.52  & 1.15 \\
981015A & 7156 & 50 - 100     & -2.6 &  430 &  6.20 & 1.54 \\
981015A & 7156 & 100 - 300   & -2.6 &  528 & 4.30  & 2.03 \\
\hline
990102A & 7293 & 20 - 50       & -2.4 &  177  & 8.52  & 1.37 \\
900102A & 7293 & 50 - 100     & -2.4 &  375 &  7.78 & 1.28 \\
990102A & 7293 & 100 - 300   & -2.4 &  566 & 5.40  & 2.02 \\
\hline
990220 & 7403 & 20 - 50       & -3.6 &  217  & 12.85  & 1.49 \\
990220 & 7403 & 50 - 100     & -3.6 &  350 &   9.34 & 1.19 \\
990220 & 7403 & 100 - 300   & -3.6 &  375 & 6.2  & 1.67 \\
\hline
\end{tabular}
\end{center}
\end{table*}

\begin{table*}
Table 1 Continued
\addtocounter{table}{-1}
\begin{center}
\begin{tabular}{| c | c | c | c | c | c | c |}
\hline\hline
GRB &  BATSE Trigger & Energy Range &         $t_o$ &  $A$ & $\tau$ & $\chi^2/\nu$  \\
\hline
990528 & 7588 & 20 - 50       & -1.3 $\pm$ 0.4 & 267 $\pm$ 19 & 4.93 $\pm$ 0.23 & 0.99 \\
990528 & 7588 & 50 - 100     & -1.3 $\pm$ 0.4 & 350 $\pm$ 25  & 4.11 $\pm$ 0.19 & 1.26 \\
990528 & 7588 & 100 - 300   & -1.3 $\pm$ 0.4 & 267 $\pm$ 39  & 3.13 $\pm$ 0.29 &  1.13 \\
\hline
990712B & 7648 & 20 - 50       & -2.2 $\pm$ 0.6 & 144 $\pm$ 21 & 7.78 $\pm$ 1.56 & 1.05 \\
990712B & 7648 & 50 - 100     & -2.2 $\pm$ 0.6 & 189 $\pm$ 28  & 7.10 $\pm$ 1.05 & 0.90 \\
990712B & 7648 & 100 - 300   & -2.2 $\pm$ 0.6 & 249 $\pm$ 18  & 5.40 $\pm$ 0.24 &  1.10 \\
\hline
000323 & 8049 & 20 - 50       & -3.4 &  249  & 14.73  & 1.17 \\
000323 & 8049 & 50 - 100     & -3.4 &  402 &  11.21 & 1.44 \\
000323 & 8049 & 100 - 300   & -3.4 &  430 & 8.15  & 2.00 \\
\hline
\end{tabular}
\end{center}
\end{table*}

In general, four channel BATSE LAD data were used.  The four channels used were channel 1 (20 - 50 KeV), channel 2 (50 - 100 KeV), channel 3 (100 - 300 KeV), and channel 4 (300 - 1000 MeV). In many GRBs, the counts in channel 4 are so low that the burst may not be easily discernible,.  Therefore, only the fits for the first three BATSE LAD energy channels are reported in Table 1. The BATSE LAD data type with 64-ms time resolution was used, as it had the highest temporal resolution that covered greatest temporal range in and around pulses.

The procedure for determining which GRB pulses were selected for fitting to the scalabale Norris pulse function was as follows.  All of the GRBs in the \citet{Nor05} catalog were considered initially. These pulses were considered initially by \citet{Nor05} because they were isolated, of relatively long duration, were relatively fluent, and had relatively long lag times between cross correlations between BATSE energy channels. A visual inspection was then done excluding some GRBs that were subjectively deemed too convolved with other pulses to yield a meaningful result.  Admittedly, this procedure might bias the results so that only pulses that might fit are selected, but this was considered acceptable since a key premise being tested is that {\it some} GRB pulses can be described by the energy-extended Norris pulse form, not all of them.

Once a GRB and its dominant pulse were selected, a range of times inside the GRB when the pulse being considered was clearly uncontaminated by surrounding pulses were recorded. The pulse light curve was then rebinned in time so that the pulse rise time occurred between about 10 to 20 time bins. This allowed the pulse to be temporally resolved while retaining a relatively large signal to noise per time bin.  Two background intervals, typically lasting 100 seconds, were selected well before the pulse and after the pulse and fit with a single third degree polynomial. A possible range of reasonable start times, $t_o$, were estimated based on visual inspection of the pulse light curves in the four BATSE LAD energy channels.

To test for goodness of fit, candidate $t_o$ values in this range were assumed, in turn. For each $t_o$ assumed, the best fit $\tau$ and $A$ values were recorded that minimized the $\chi^2$ per degree of freedom ($\nu$, the number of time bins) for each energy channel. The value of $t_o$ that allowed the best fit $A$ and $\tau$ values were recorded, as well as error estimates for these values.

There are many factors that can affect the statistics of the fit so as to render the computed goodness of fit values more indicative than numerically definitive. These include the fineness of the time binning, the precision of the background fit, the temporal range of pulse bins fit, and the affect of secondary independent pulses. In general, breaking the pulse up into more time bins tended to lower $\chi^2/\nu$, although the increase in $\nu$ then tended to offset significant changes in actual goodness of fit. Although pulse fits always started at $t_o$, extending the temporal fit range of the pulse decay further into the background tended to drive $\chi^2/\nu$ down since the more slowly changing background was typically easier to fit.  The pulse was deemed to have ended, in terms of the statistical fitting procedure, when the pulse was either a few percent above background or a second contaminating pulse became clearly apparent. Given these boundaries and limitations, a pulse fit was considered descriptively useful when $\chi^2/\nu$ was computed to be 1.5 or lower over the fitted pulse range.

\begin{figure*}
\includegraphics[scale=0.30, angle=90]{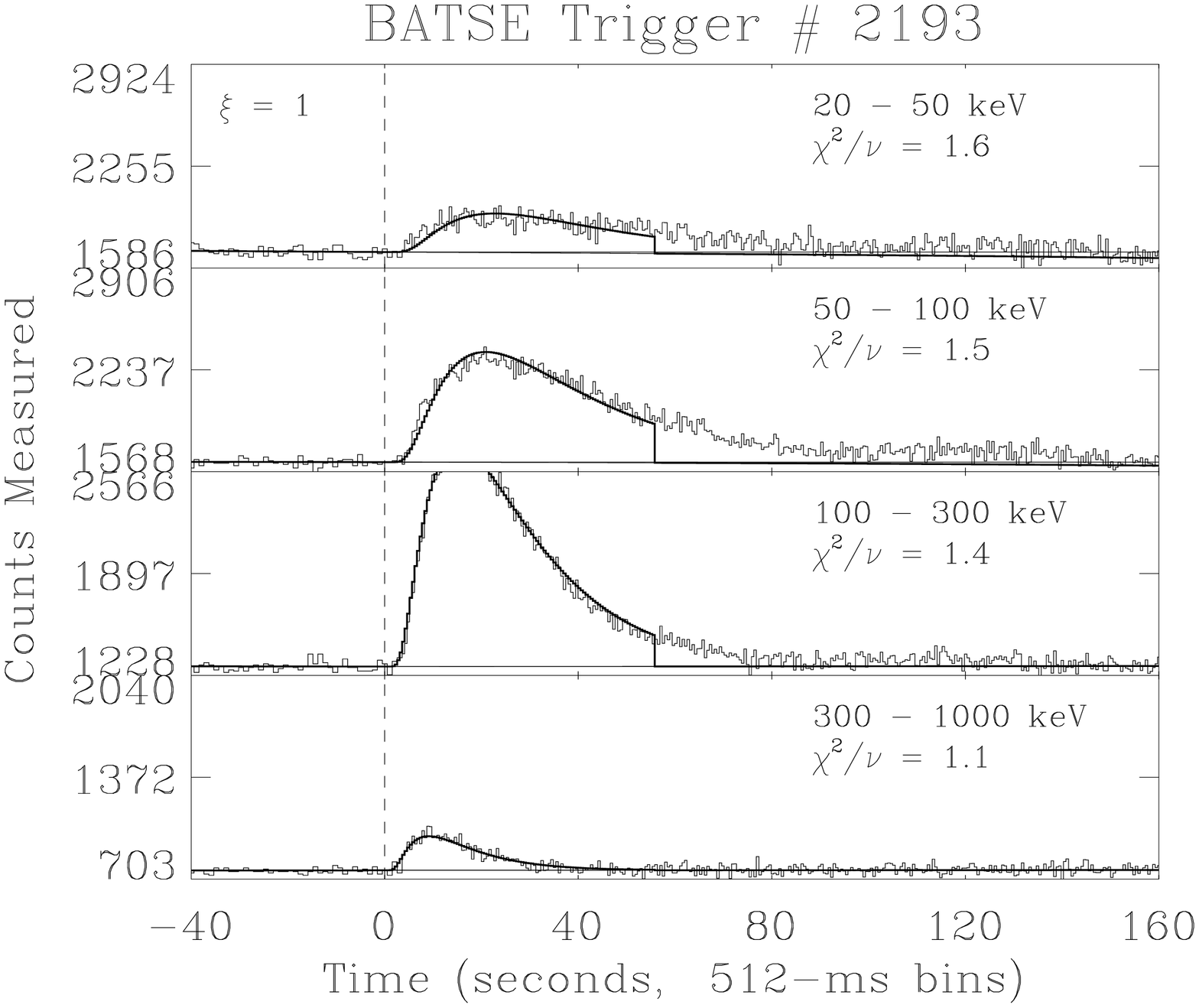}
\includegraphics[scale=0.30, angle=90]{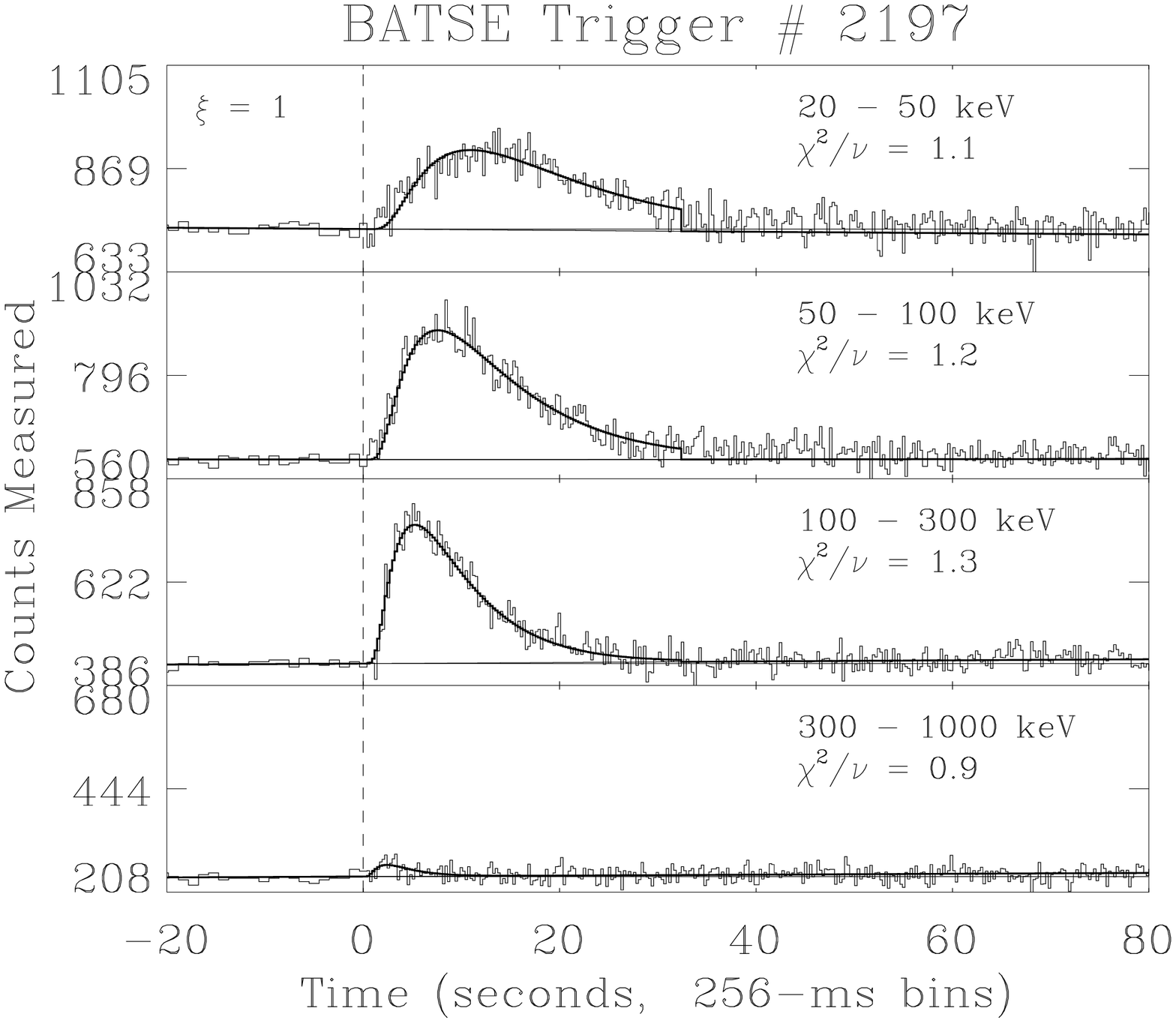}
\includegraphics[scale=0.30, angle=90]{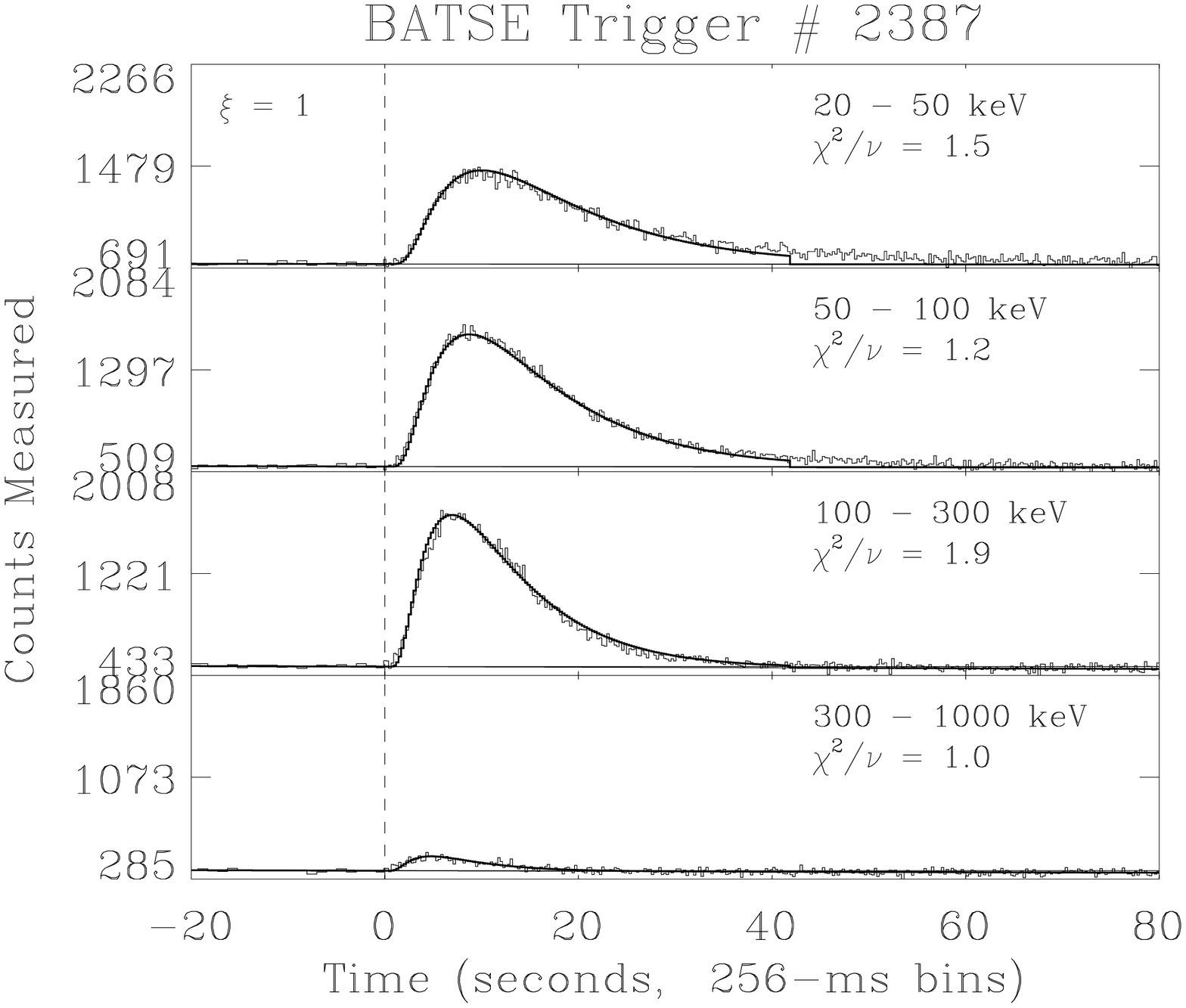}
\includegraphics[scale=0.30, angle=90]{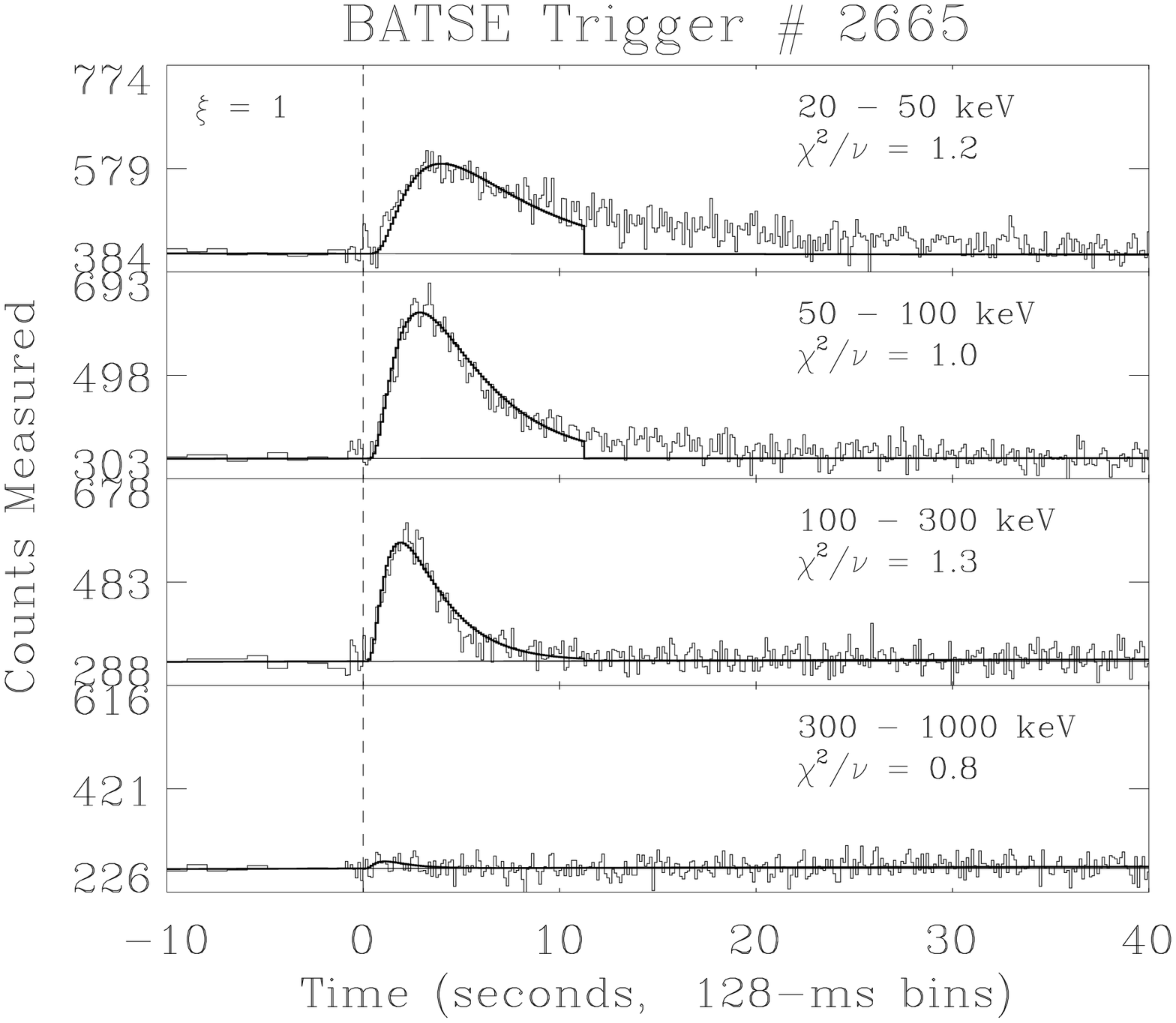}
\includegraphics[scale=0.30, angle=90]{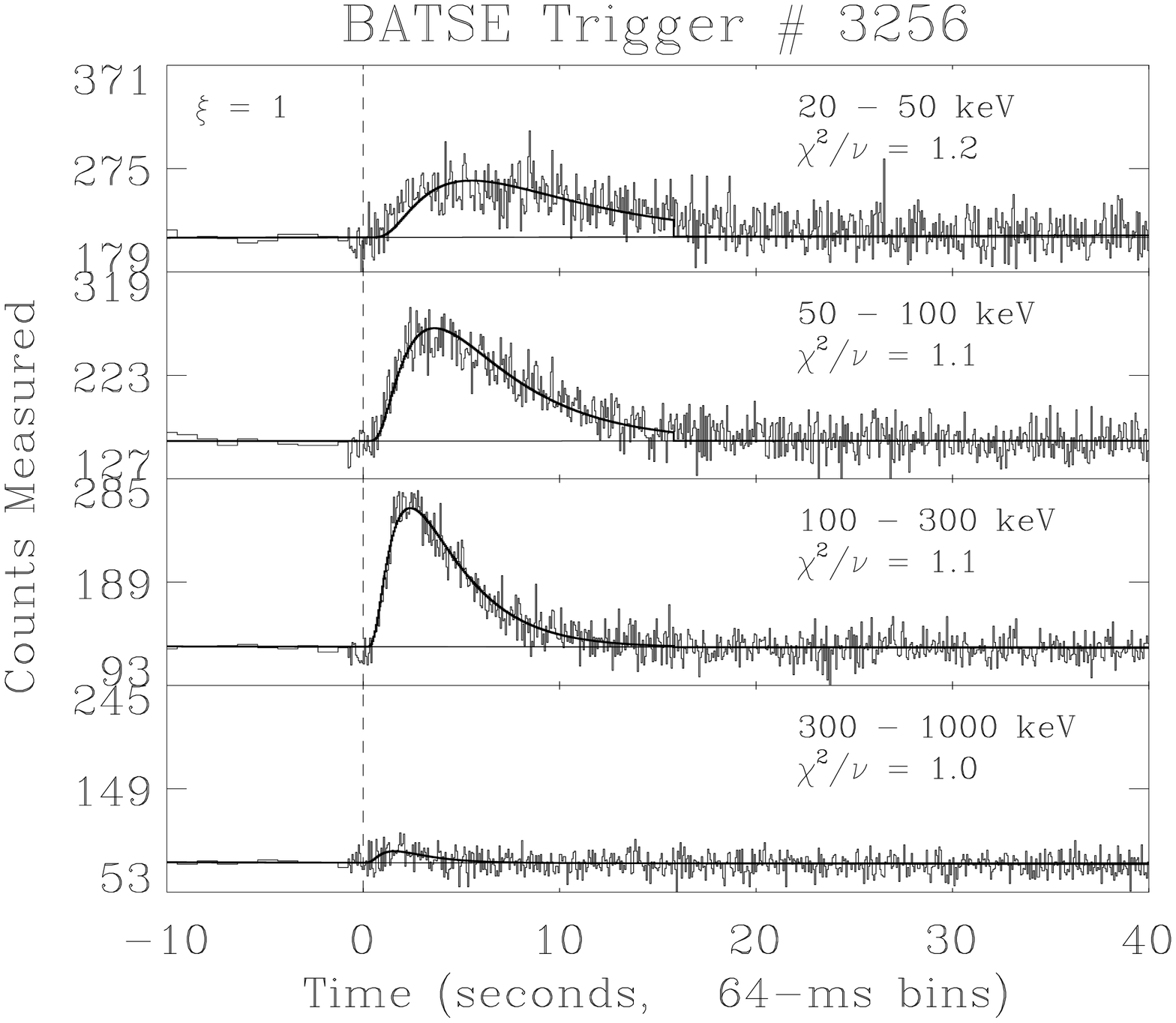}
\includegraphics[scale=0.30, angle=90]{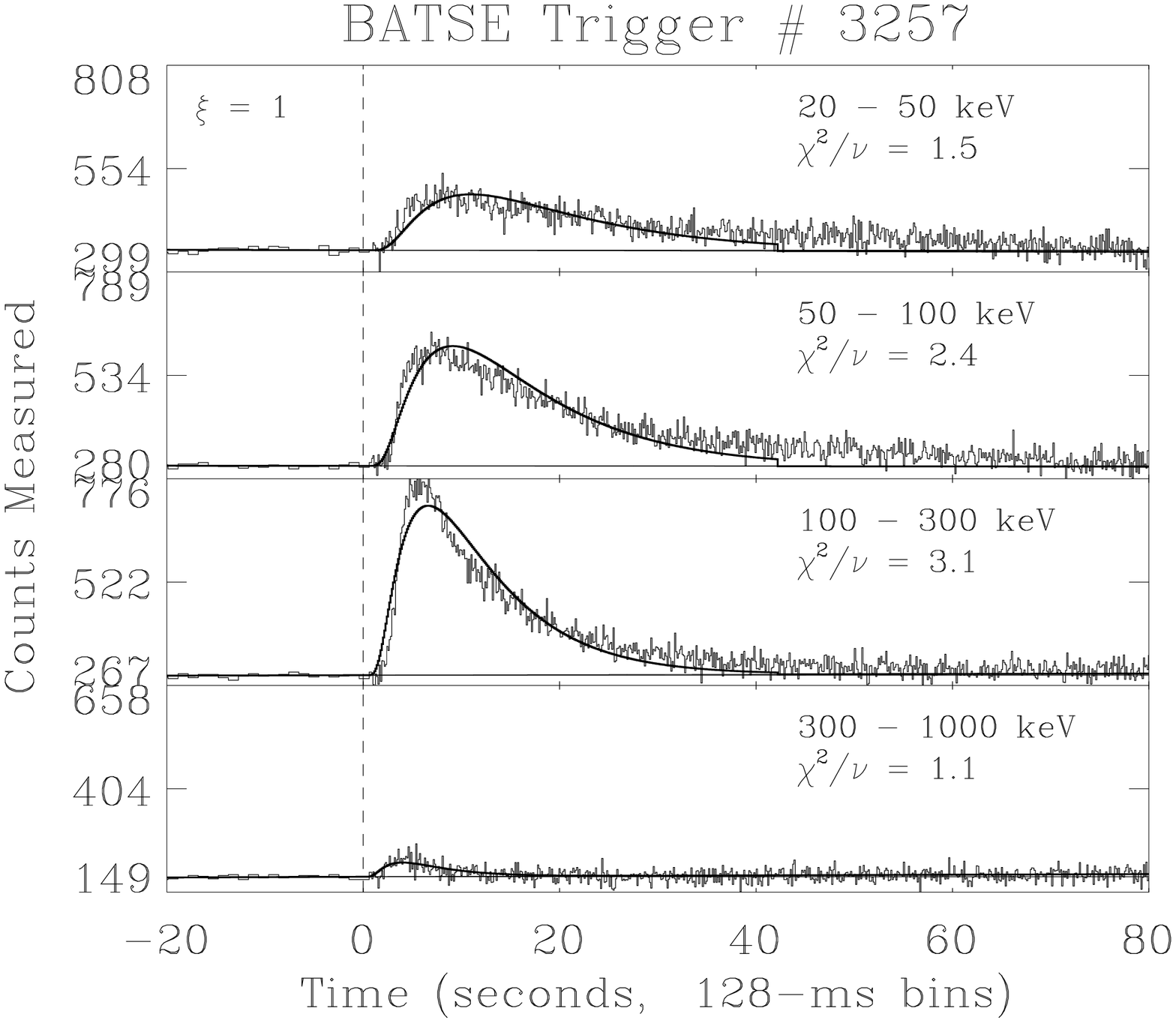}
\includegraphics[scale=0.30, angle=90]{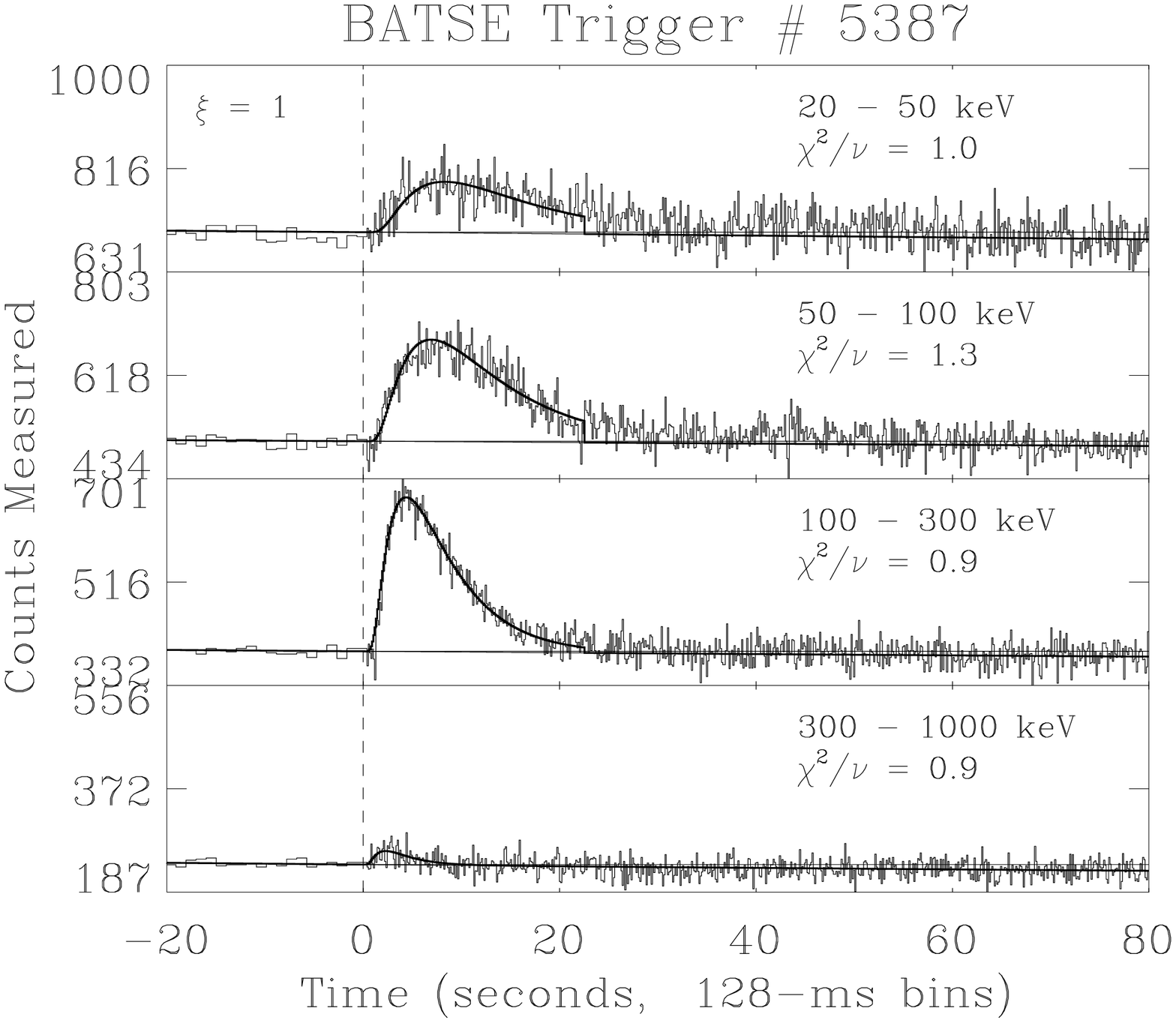}
\includegraphics[scale=0.30, angle=90]{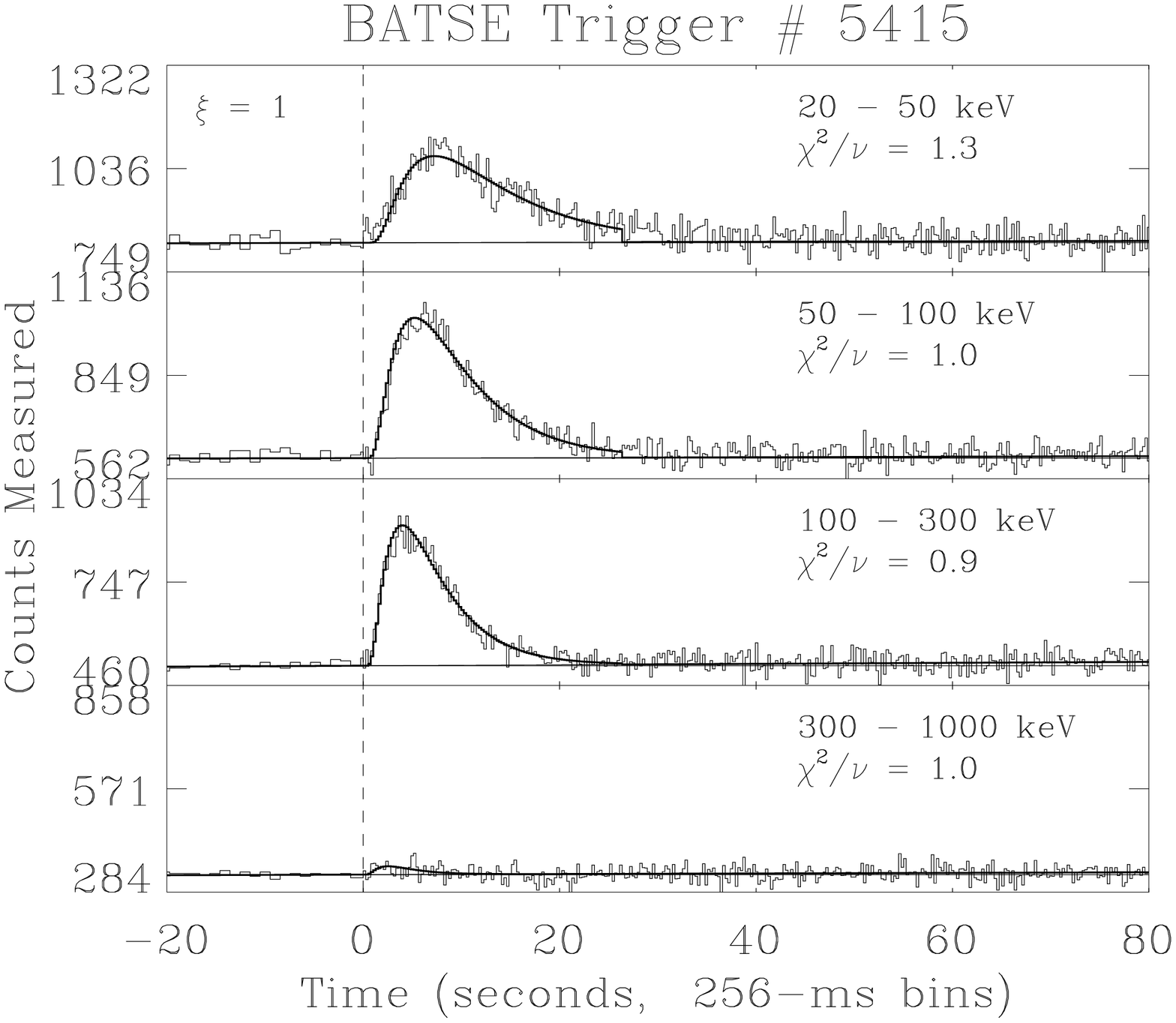}
\caption{Plots of GRB light curves.}
\label{fitplots}
\end{figure*}

\begin{figure*}
\includegraphics[scale=0.30, angle=90]{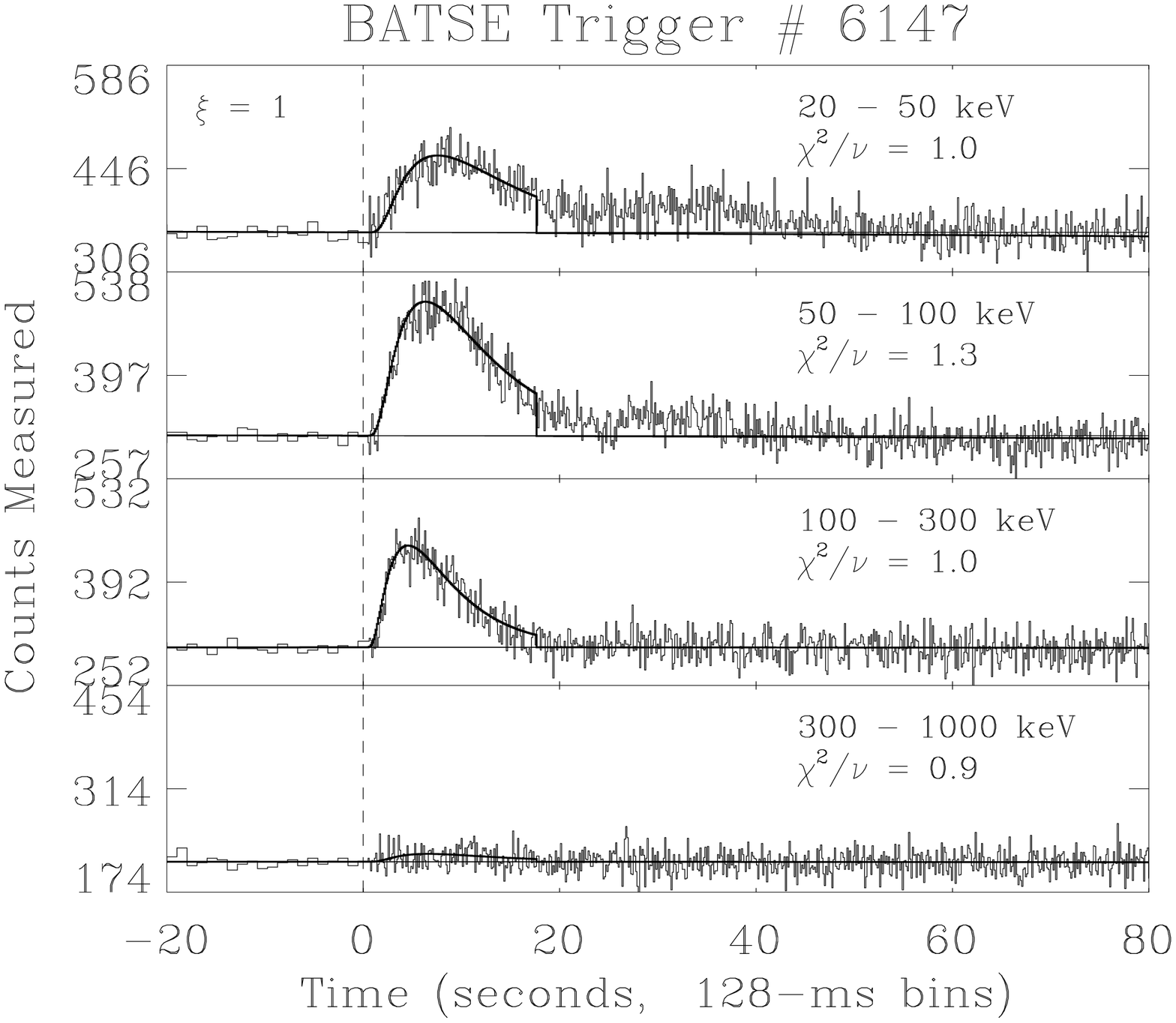}
\includegraphics[scale=0.30, angle=90]{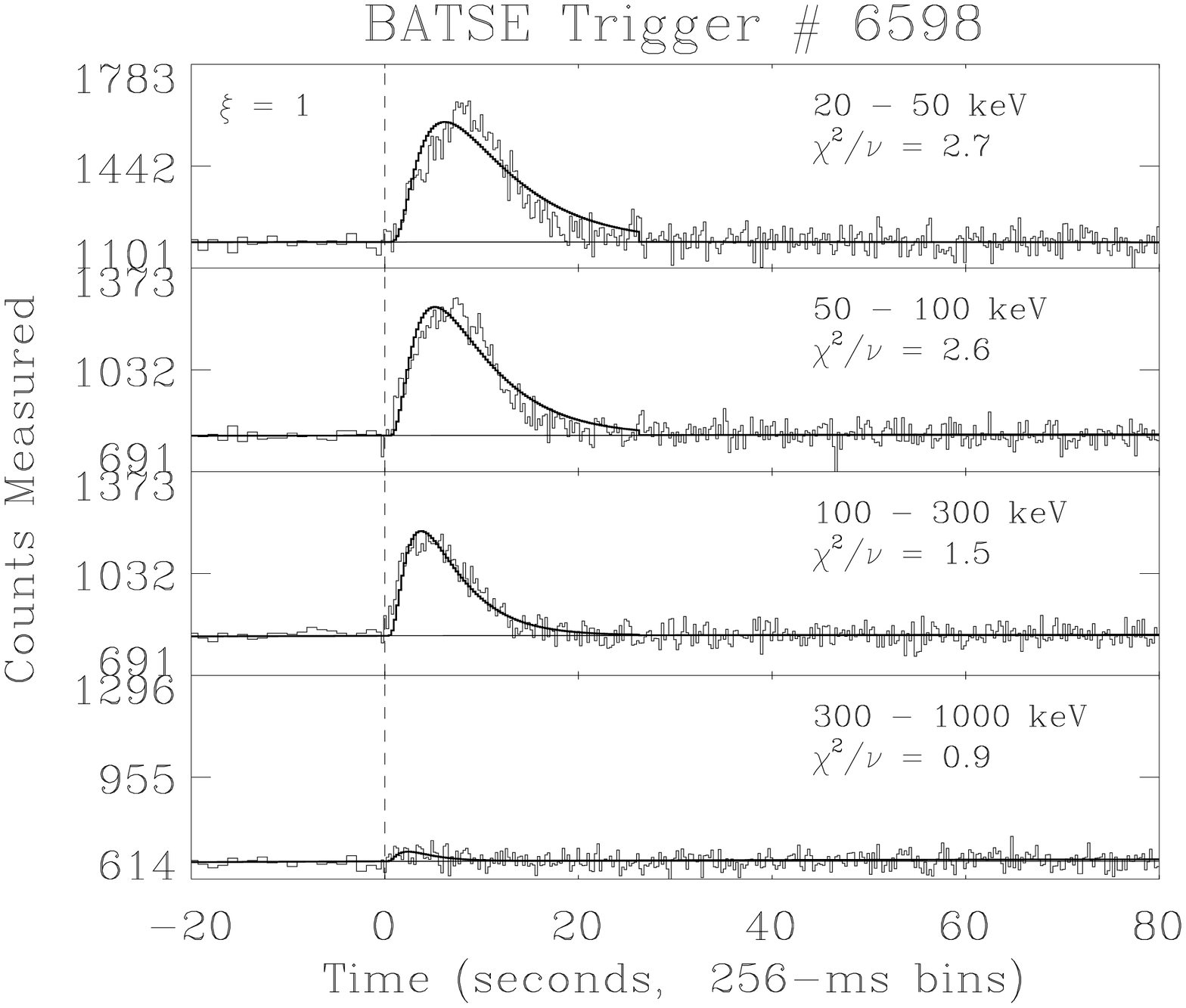}
\includegraphics[scale=0.30, angle=90]{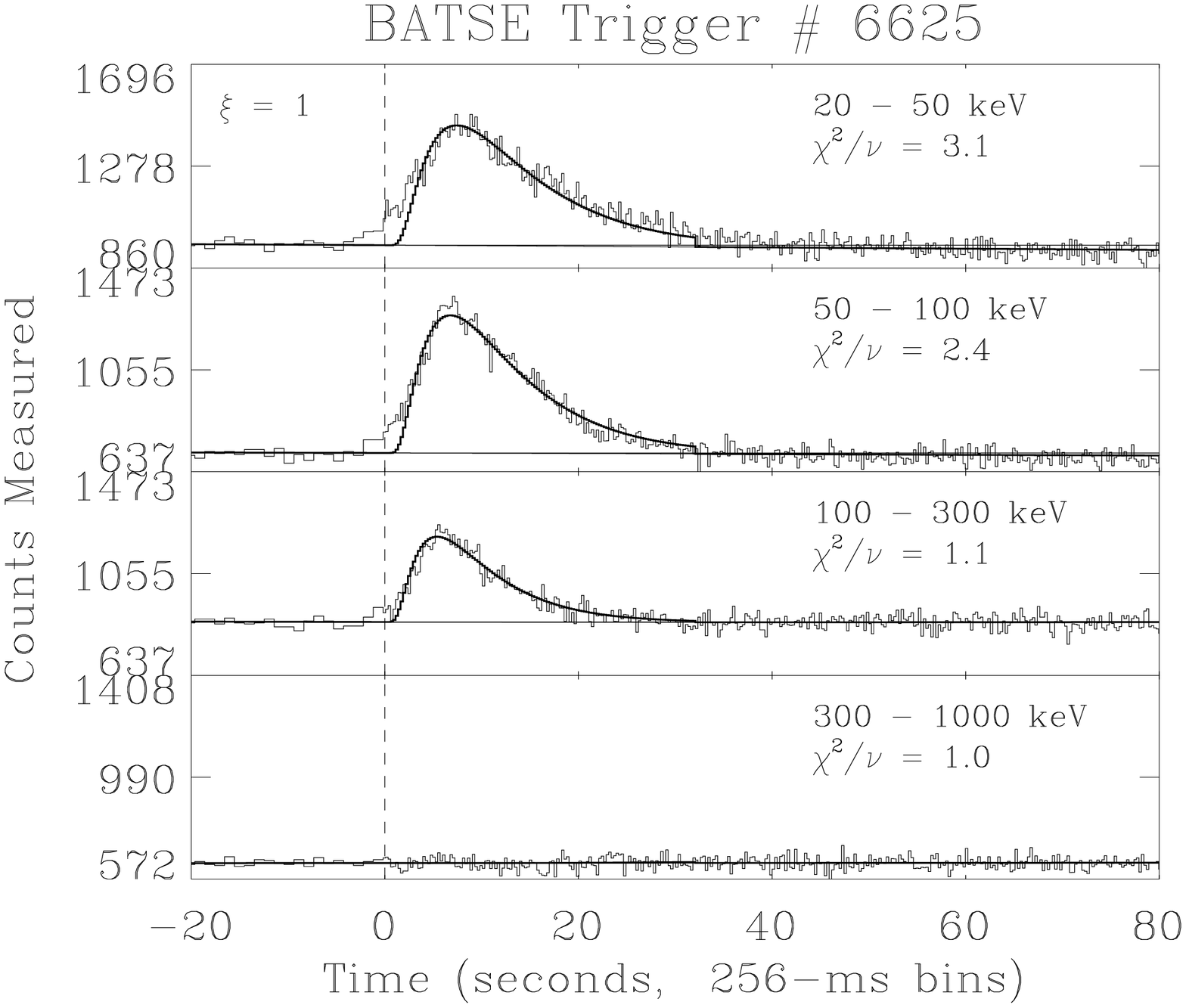}
\includegraphics[scale=0.30, angle=90]{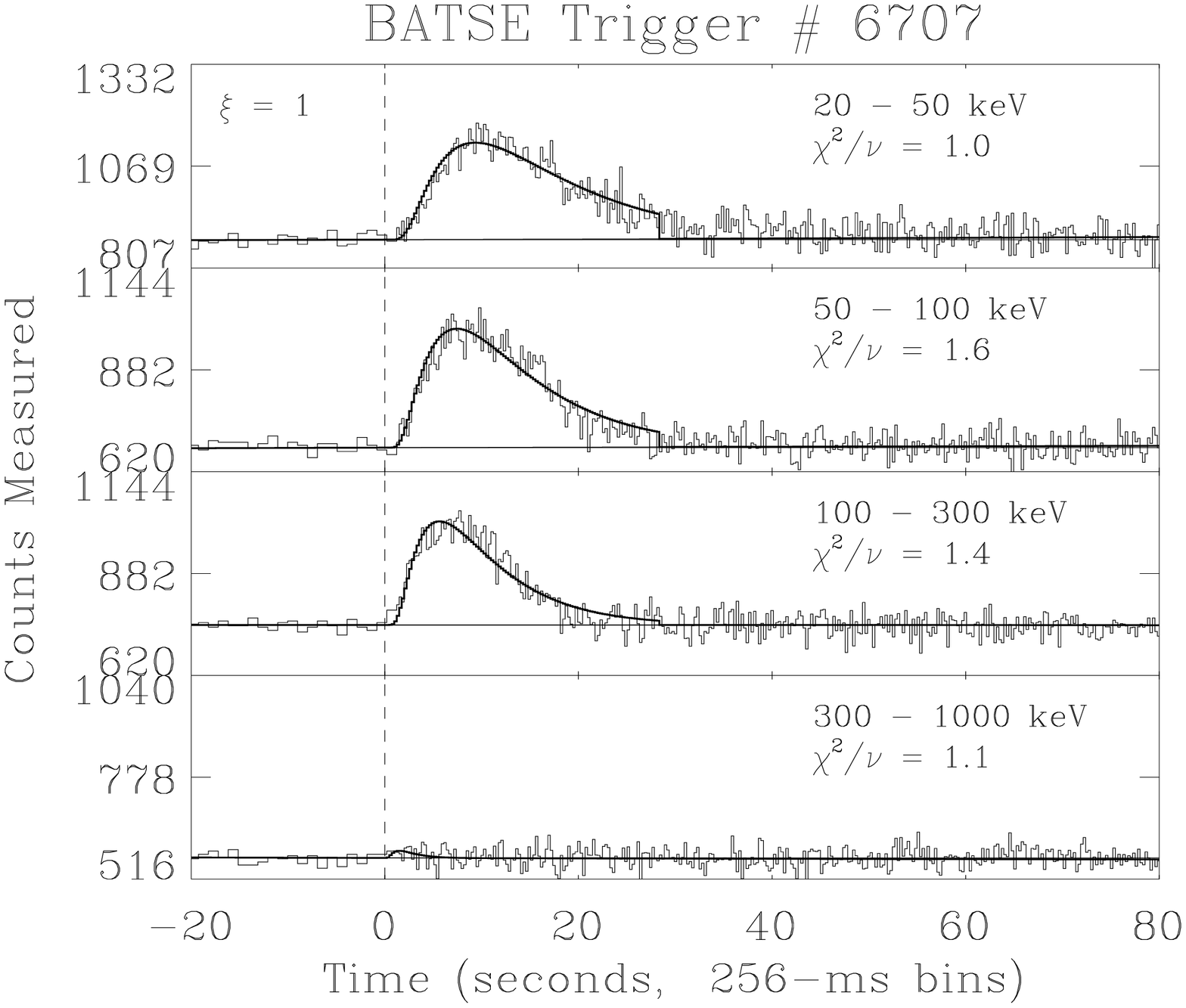}
\includegraphics[scale=0.30, angle=90]{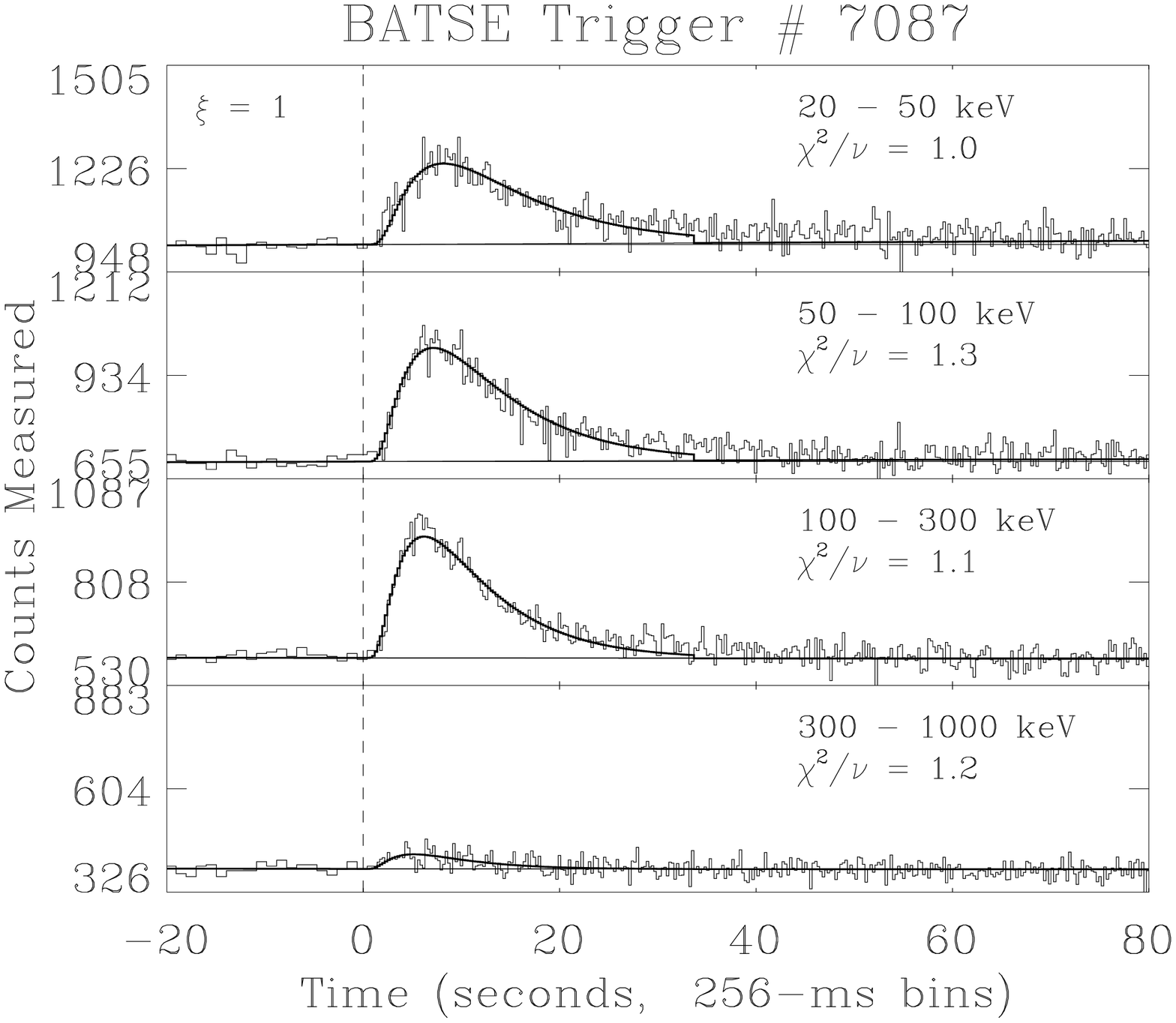}
\includegraphics[scale=0.30, angle=90]{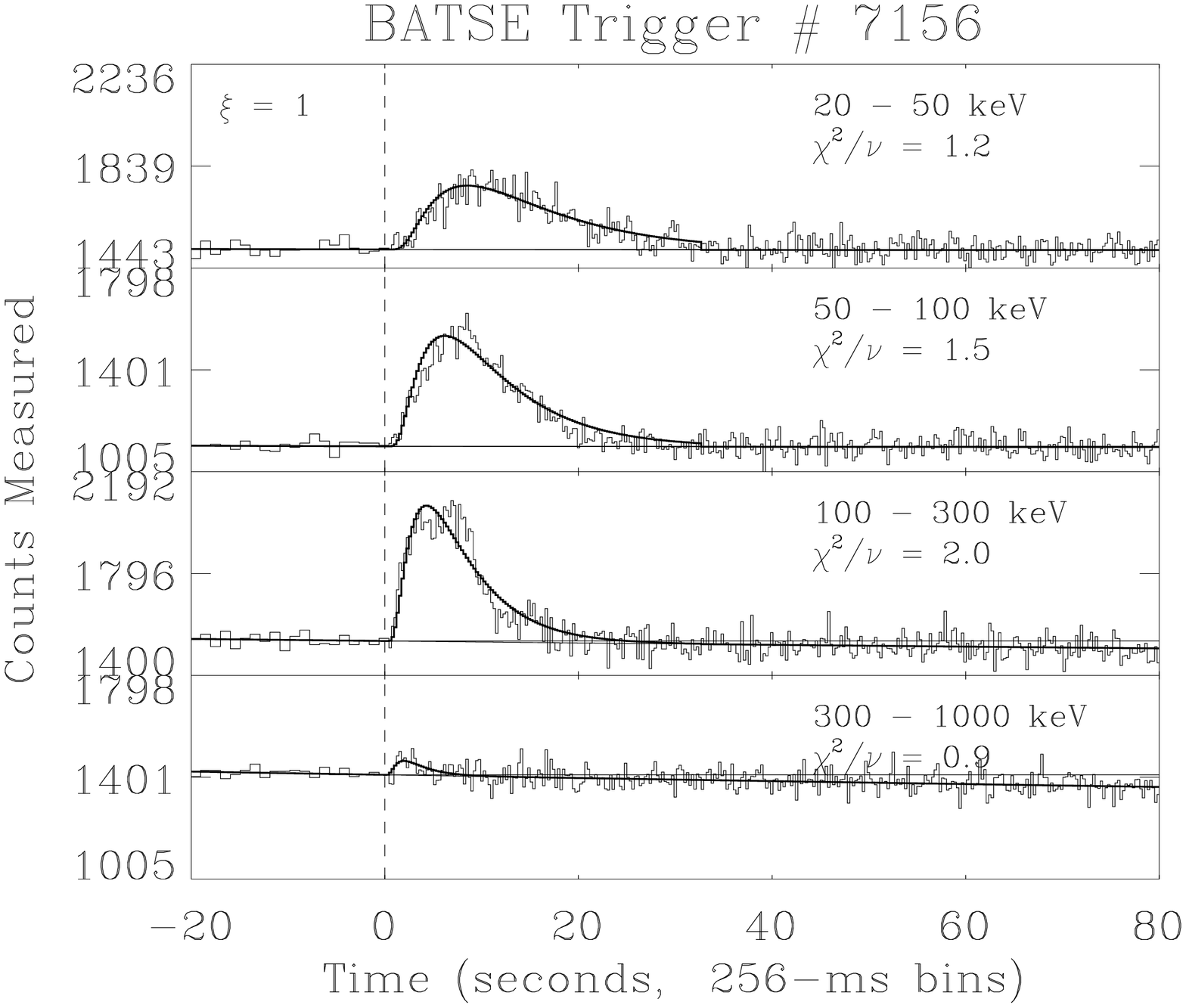}
\includegraphics[scale=0.30, angle=90]{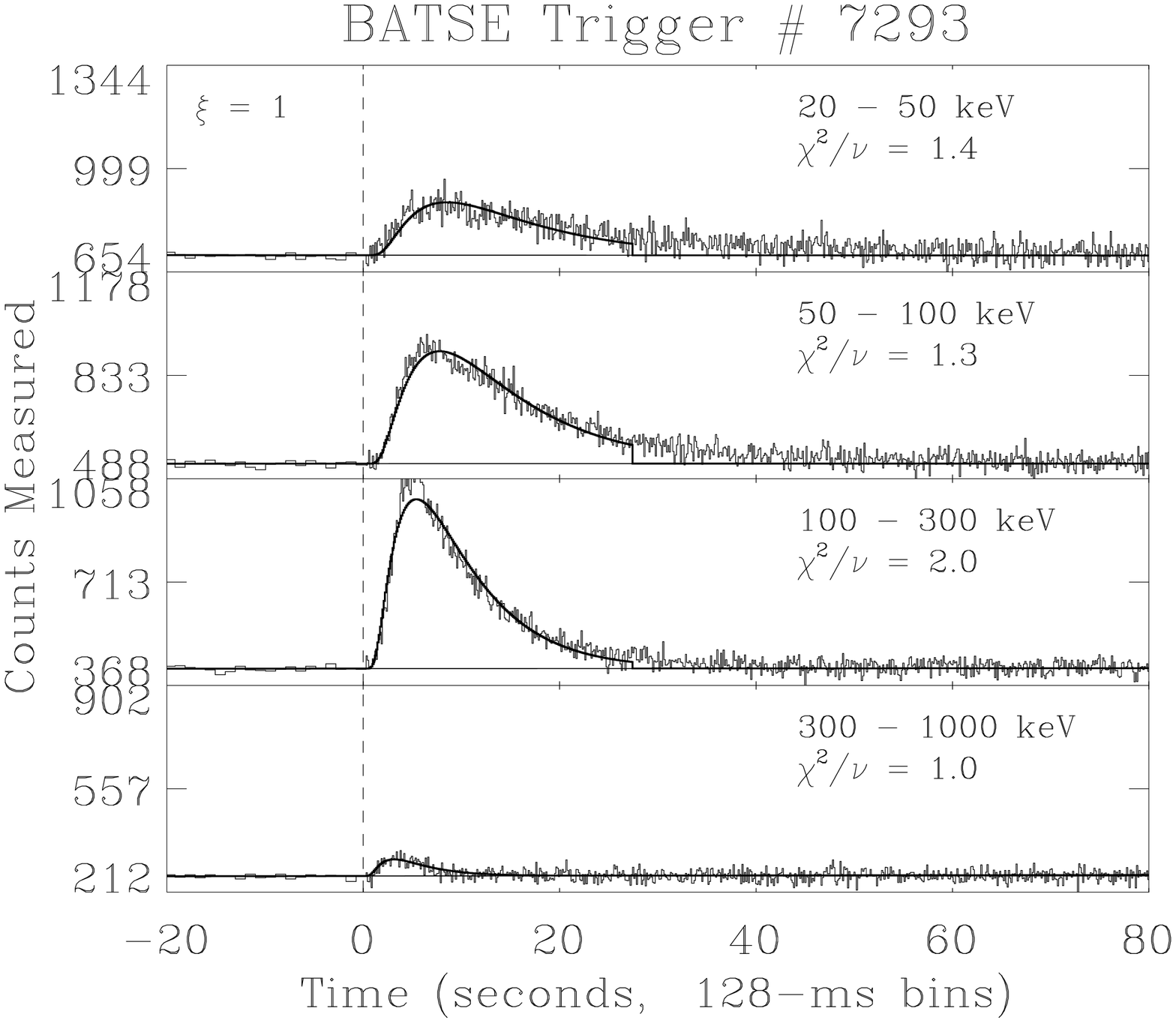}
\includegraphics[scale=0.30, angle=90]{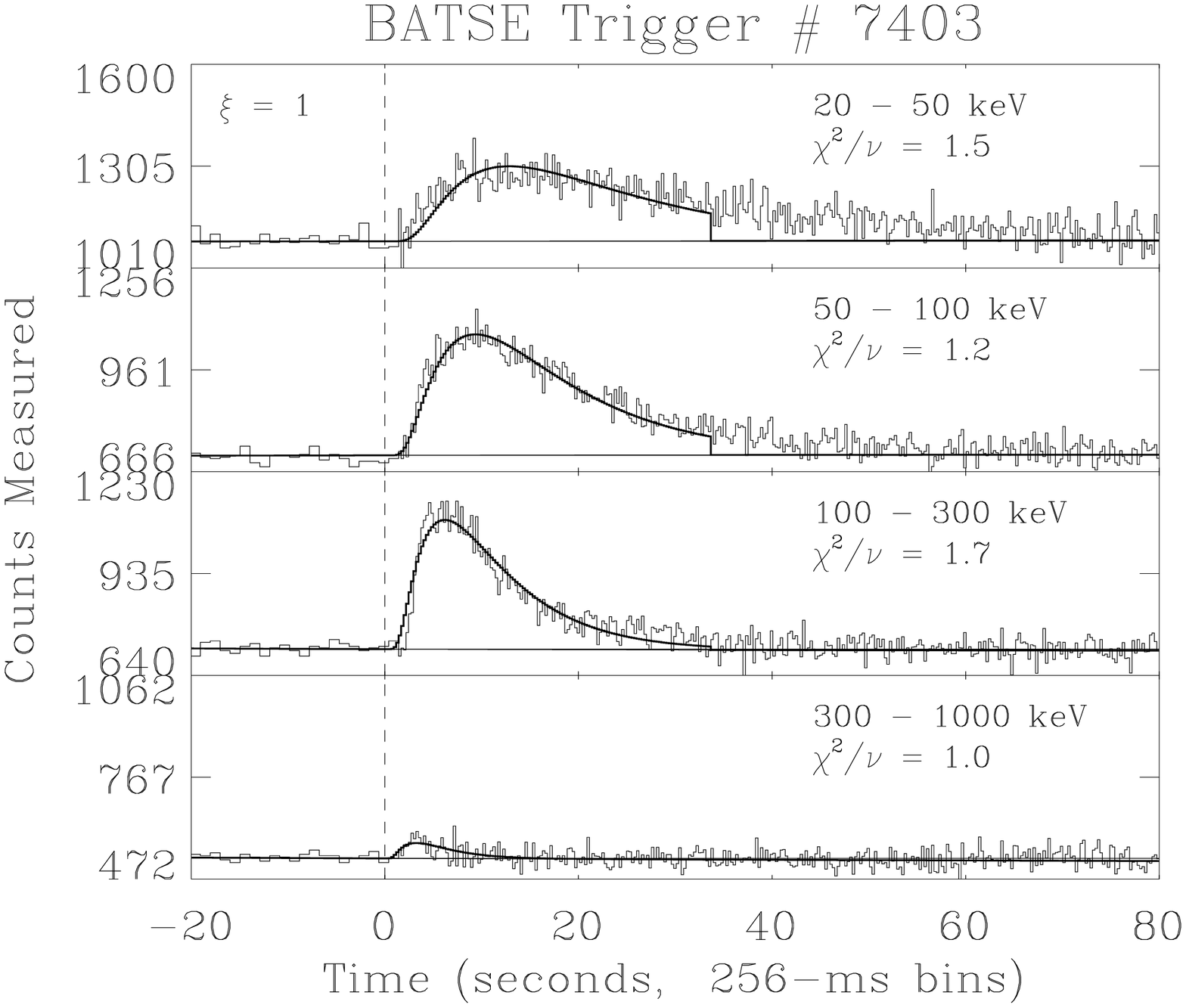}
\addtocounter{figure}{-1}
\caption{Continued}
\end{figure*}

\begin{figure*}

\includegraphics[scale=0.30, angle=90]{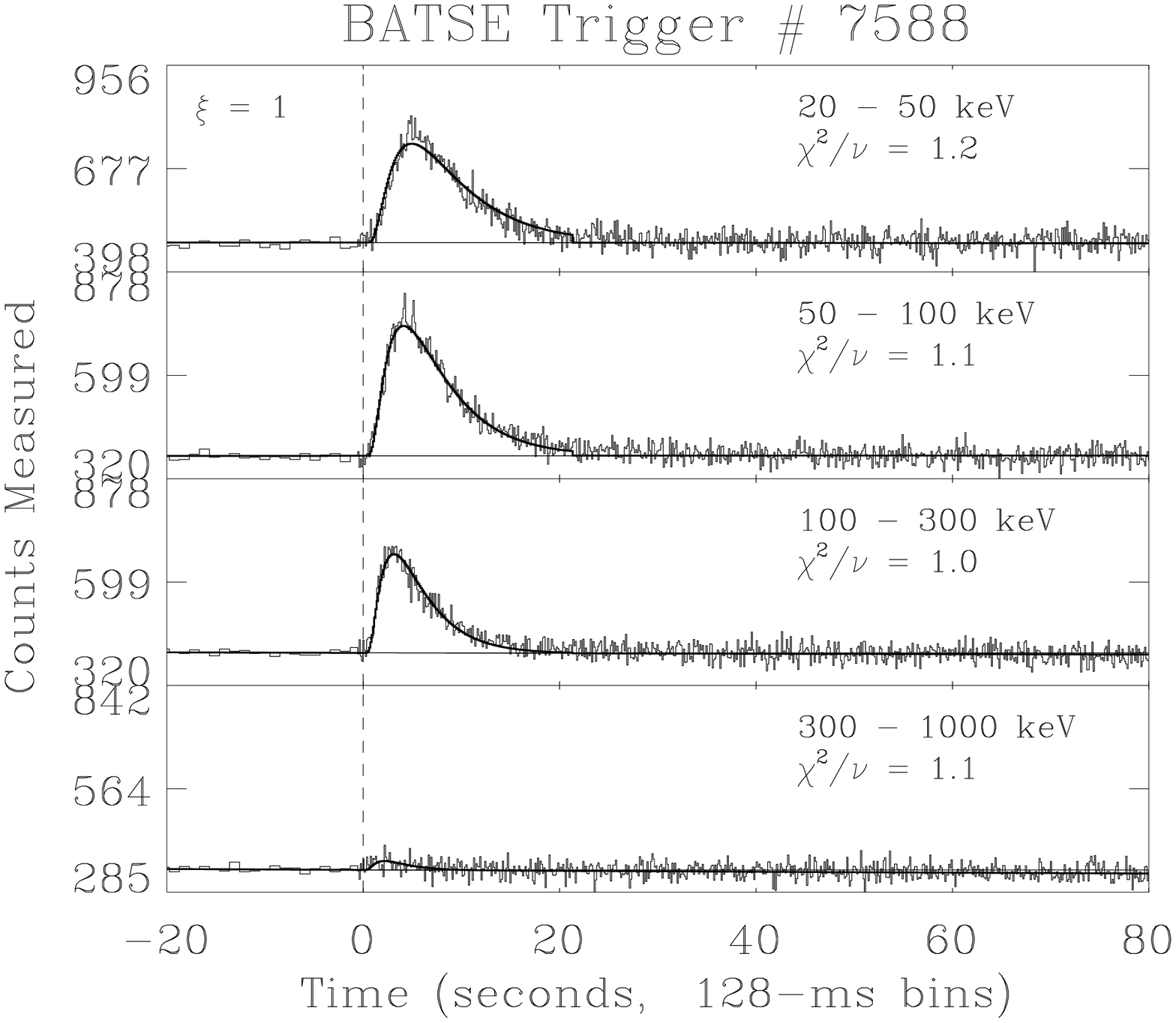}
\includegraphics[scale=0.30, angle=90]{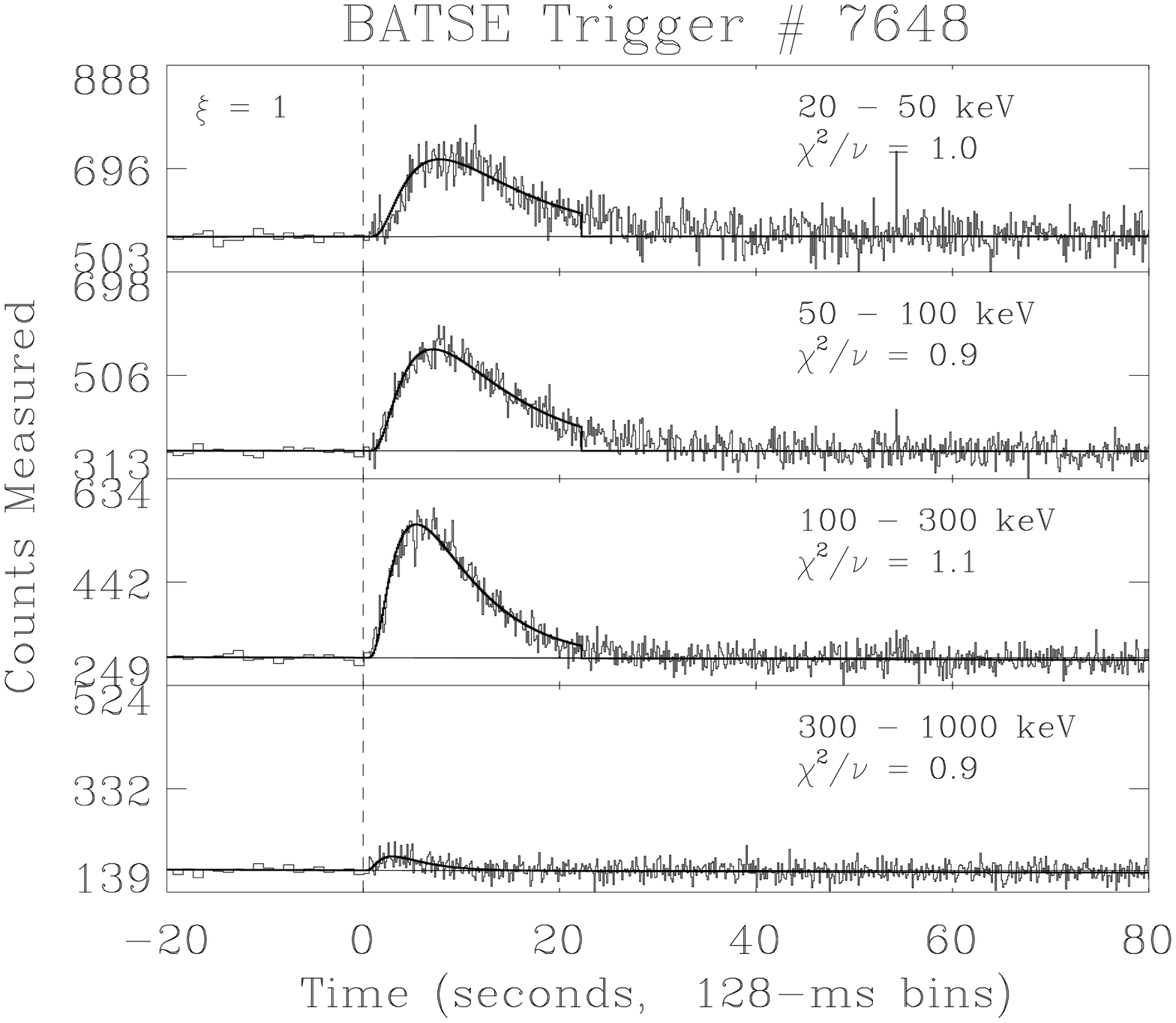}
\includegraphics[scale=0.30, angle=90]{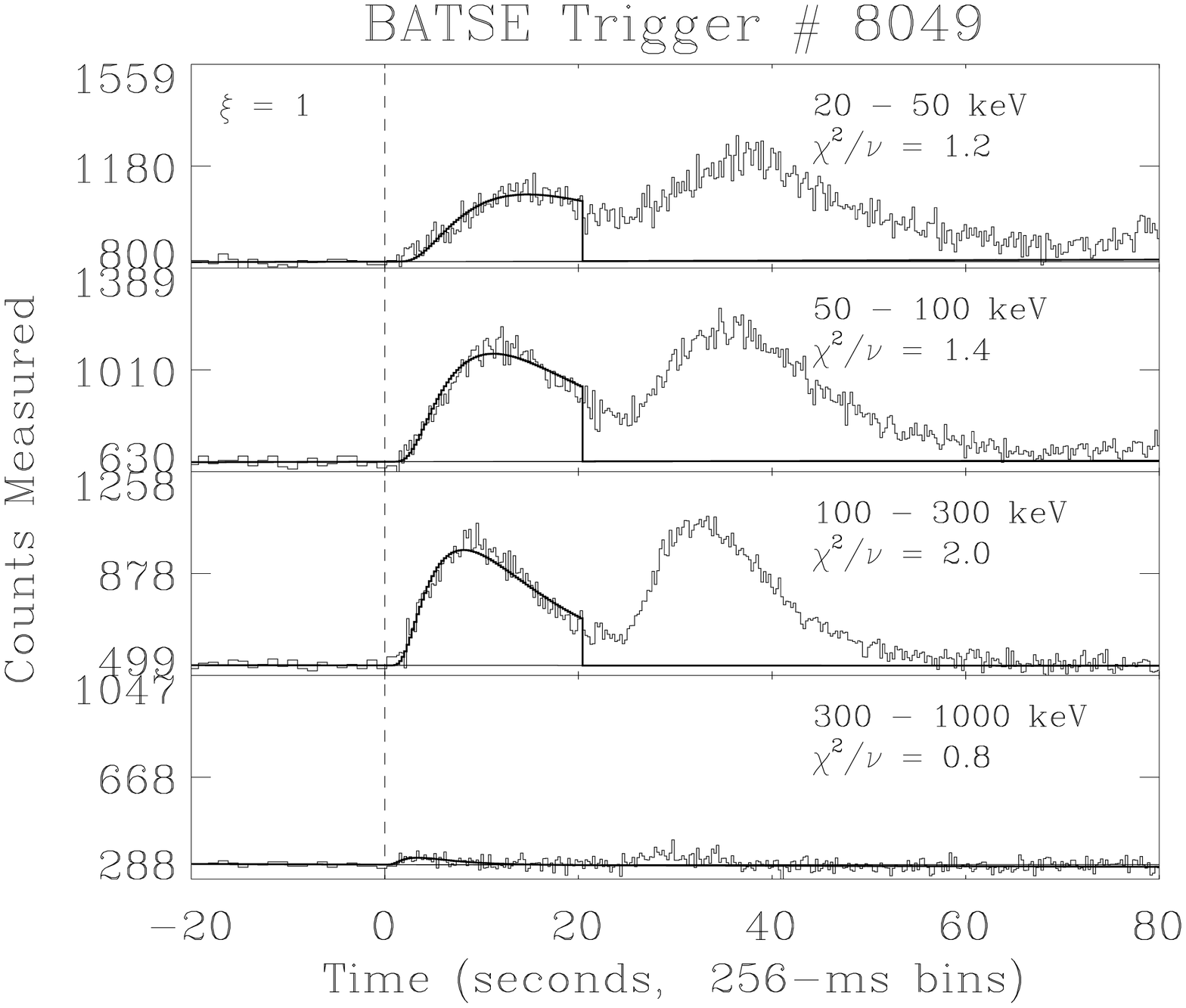}
\addtocounter{figure}{-1}
\caption{Continued}
\end{figure*}

The results of the fitting process are shown in Figure (\ref{fitplots}) with the best fitting parameters listed in Table 1.  For a GRB's considered pulse, a best fit was declared across energy channels when the sum of the squares of $\chi^2 / \nu$ over all of the energy channels was at a minimum. In Figure (\ref{fitplots}), light curves for the GRB are given for the lowest three BATSE energy bands, with the lowest energy given at the top. The raw BATSE data is shown in histogram form by the fluctuating solid line. Time is depicted relative to the start time of the dominant pulse shown, with a vertical dashed line making this start time $t_o$ more clearly. Superposed on the data are the best fit plots for the Norris pulse model expanded across energy bands as described above.

In Table 1, the first column lists the GRB date designation, while the BATSE trigger number is given in column 2, and the BATSE energy range being fit is listed column 3.  The best fit start time $t_o$ relative to the trigger time for the dominant pulse in the GRB is given in column 4, along with an error estimate.  For this start time, the best fit parameters for $A(E)$ and $\tau(E)$ are given in columns 5 and 6 respectively, along with their error estimates. The $\chi^2 / \nu$ goodness of fits values are listed in column 7.

Error estimates are only given for GRBs where the$\chi^2 / \nu$ was below 1.5 for every energy channel. For GRB pulses where the fit was deemed ``not good," meaning here that at least one energy channel could not be fit with the minimal Norris pulse function to a $\chi^2/\nu$ better than 1.5, then only the best fit $t_o$, $A$, and $\tau$ parameters are listed.  Where a good fit was found, the $t_o$, $A$, and $\tau$ values, as well as their errors, are listed. The errors indicate the limiting value of the parameter that could create a good fit, given the best fit listed for the other parameters.

For BATSE 6598, a better fit was obtained for $\xi > 1$. Specifically $\xi = 2$ provided a better fit in almost all energy channels.  BATSE trigger 6625 appears upon inspection to contain an early pulse that is biasing the statistics.  In BATSE trigger 8049, only the first pulse was fit.  This is an example of fitting a pulse based on incomplete information, which this technique allows.  Note that for BATSE 8049, the second pulse was not used since one would first have to model the first pulse and subtract it, which was considered beyond the scope of this work.  Given the Pulse Start Conjecture, though, the second pulse is hypothesized to be completely contained after a time that can be estimated by inspection of the light curve.

\section{A Search for Correlations Between Scalable Pulse Parameters}

From inspection of Table 1 and the plots in Figure {\label{fitplots}), it is clear that a group of GRBs exist that contain a pulse that statistically satisfies all three of the above stated pulse scale conjectures simultaneously.  From the GRBs tested here, this group includes BATSE trigger numbers 2197, 2665, 3256, 5387, 5415, 6147, 7087, 7588, and 7648.  The dominant pulse of these well-fit GRBs will be called an Energy Scalable Pulse (ESP).

Inspection of the $A$ values in Table 1 shows that the full pulse spectra for ESPs can be quite diverse.  Typically, a spectrum will peak in one of middle BATSE energy bands, hence showing an $E_{peak}$ value common to the range of $E_{peak}$ values in published GRB spectra.  The existence of $E_{peak}$ in this range might not be coincidence and might be related to the trigger criteria of BATSE GRBs.  The low energy resolution of the spectra make it difficult to determine if the full pulse spectra follow a Band functional form or any other spectral form common to other types of GRBs.

\begin{figure}
\includegraphics[width=84mm]{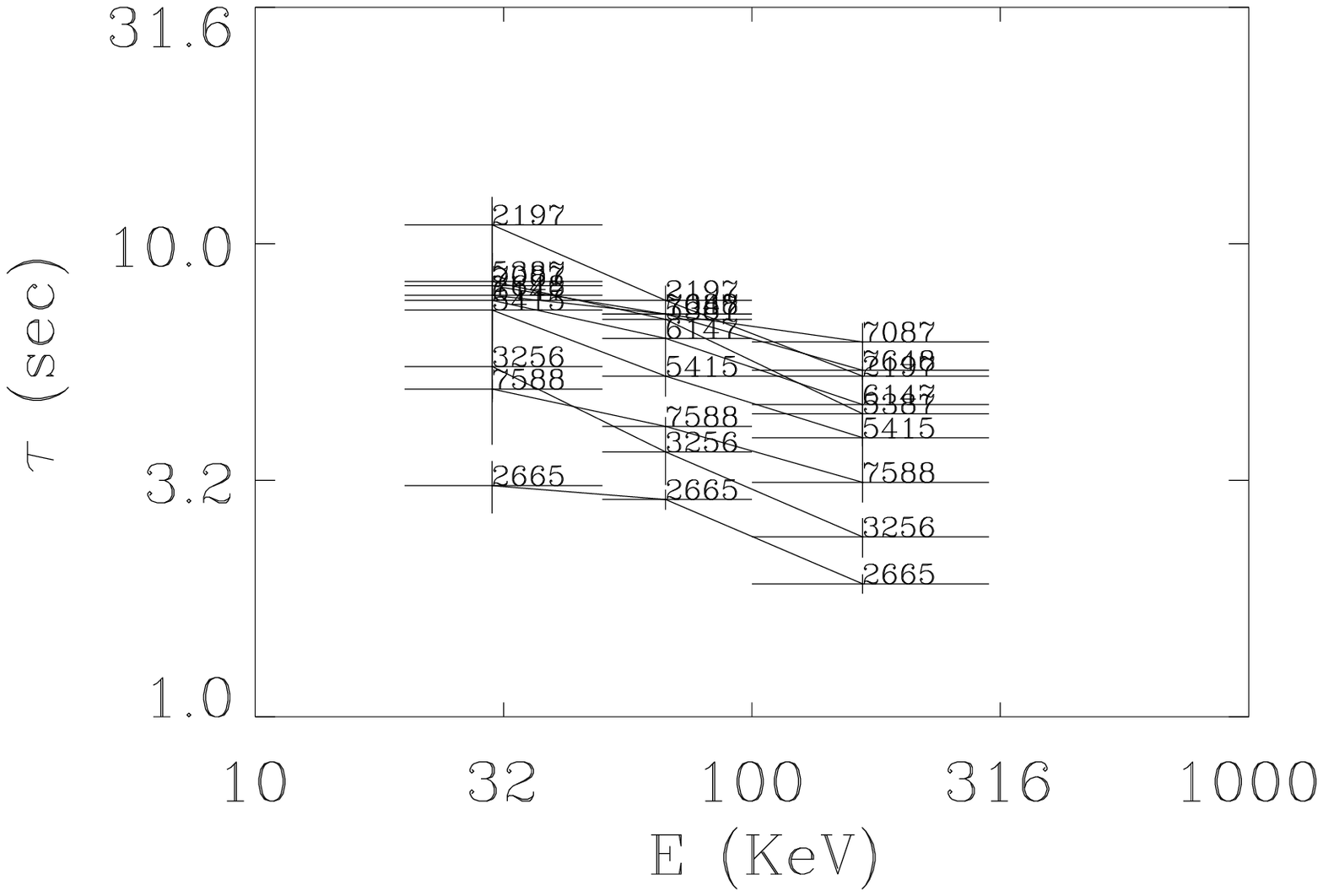}
\caption{A plot of pulse stretching factor $\tau$ versus pulse energy $E$ for selected pulses where where the Norris pulse model approximations provided reasonably good data fits to BATSE data.  Error bars are included, where the error in $E$ was just the width of the BATSE energy channel. Each point is labeled by its BATSE trigger number.  Note that for most GRB pulses fit, $\tau$ may be related to $E$ by a power law.}
\label{tautrend}
\end{figure}

A plot depicting how the  temporal stretch factor $\tau$ changes with energy for the ESP sample is shown in Figure (\ref{tautrend}).  Each GRB is labelled by its BATSE trigger number.  Note that $\tau$ is usually consistent with a power law in energy to within the uncertainty.  There is at least one exception to this rule, in that BATSE trigger 2665 does not appear to have a power law relation between $\tau$ and $E$.

Furthermore, the exponent of the power law relationship between $E$ and $\tau$ does not appear to be consistent between GRBs.  For example, the dominant pulse in BATSE trigger 2197 appears well fit by $\tau \propto E^{-0.5}$, yielding a relatively large difference between the duration of the pulse at high and low eneries.  In contrast, the dominant pulse in BATSE trigger 7087 appears well fit by $\tau \propto E^{-0.17}$, yielding a relatively small difference between the duration of the pulse at high and low energies.

Note that the difference between the $\tau$ factors of two energy channels is conceptually similar to the lag between these two energy channels.  It is therefore possible that the power law exponent between $\tau$ and $E$ might act similar to pulse lag and be a measure of intrinsic brightness of the pulse and hence the GRB. Since $\tau$ is the time between the pulse peak and the pulse start, the difference between two $\tau$s of the same pulse at different energies is just the difference between the peak times of the pulse at those energies.  Although the lag is formally computed by noting the time of the maximum in the cross correlation between the light curves of two energy channels, this cross correlation would have it's greatest instantaneous power when the two peaks are aligned. Therefore lag is strongly pulled toward the difference in peak times. Therefore, one would expect ${\rm lag}_{23} \sim (\tau_3 - \tau_1)$.

\section{Discussion and Conclusions}

It is evident from inspection that GRB pulses have similarities at different energies.  Here clear rules are postulated for transferring specific pieces of information between energies.   Specific analyses presented above indicate that at least one class of GRB pulses, here dubbed ``energy scalable pulses" (ESPs), can be described by a simple, cross-energy mathematical form that has a well-defined start time and a common light curve shape that differs between energies by only scale factors is time and brightness.  The particular light curve shape explored here is the simple one given in \citet{Nor05}: $P(t) = A e^{-(t/\tau + \tau/t)}$ where $t$ is time since an energy independent pulse start $t_o$, and (only) $A$ and $\tau$ are functions of energy. Other simple light curves descriptions are also possible \citep{Nem11}.

Implications of the above results can be broken into two categories: computational and physical.  Computationally, ESPs can be adequately fit to real data by exploring a reduced parameter space when compared to non-ESPs. For example, fitting a non-ESP, or ignoring the ESP nature of a pulse, may cause it to be fit by four free parameters per pulse per energy channel: $t_o$, $A$, $\tau$, and $\xi$.  Invoking the ESP model halves the number of fit parameters to explore, since start time $t_o$ and pulse shape $\xi$ are postulated to be the same at each energy.  The time saved computationally may exceed a factor of two since the number of nested loops is halved once $t_o$ and $\xi$ have been determined. Furthermore, $\xi$ may not even have to be determined when postulated to be same for different pulses.

The physical implications of the existence of ESP pulses might be numerous, but only a few aspects will be touched on here.  First, the scalable nature of complete ESPs between energies demands that the rise and fall of the pulse is dependent on the same temporal scale factor $\tau$, perhaps indicating a common energy liberation mechanism \citep{Nem00, Hak11}.  This is counter to analyses in \citet{Nor05} where the pulse rise and fall are considered separate, implying that the energy liberation mechanisms might also be separate.  It is still possible, for example, that a correlation exists between duration $\tau$ and pulse shape $\xi$ for non-ESP pulses \citep{Pen10, Hak11}.

Next, the common start time $t_o$ found here for ESPs across energy bands further bolsters the case that GRB pulses are generated by some physical triggering mechanism related to that start time.  Whether $t_o$ marks the time of an explosion, a collision, a phase change, or something completely different, is not speculated on here.

It is interesting that the error in this start time is typically much less than the duration of the pulse, typically by at least one order of magnitude.  In the case of the dominant pulse in BATSE GRB 5415, the ratio of $\Delta t_o / \tau$ was recorded to be about 0.055 for energy channel 1, while $\Delta t_o$ was estimated to be about 0.4 sec.  Seemingly trivial results like this might be of particular interest because they might provide important limits on any dispersive property of the universe between us and cosmologically distant GRBs \citep{Nem12}.

Next, the common shape parameter $\xi=1$ shared by many ESPs may indicate that a single set of physical parameters may be operating on these pulses.  Physical mechanisms that are capable of generating a shape statistically similar to the Norris pulse shape should be explored.

Why aren't all GRB pulses ESP pulses?  It is possible that they are, but that the temporal overlap between multiple pulses is hiding this attribute for most pulses.  In many of the cases of \citet{Nor05} selected pulses, a dominant pulse might be contaminated by at least one much smaller amplitude pulse.  One possible example of this is BATSE trigger 6625, shown in Figure 4.  There, most of the late time behavior of the pulse appears well fit by the ESP paradigm, while a faint pulse may contaminate the start of the rise period of the dominant pulse.

Possibly, whatever processes that create GRB pulses actually generate a whole luminosity function of pulses with lower amplitudes and lower energies being more common. Then, most detected GRBs are such dense conglomerations of these faint and weak pulses that it is not possible to separate out even one pulse for detailed analysis.  The selection of the \citet{Nor05} sample was an attempt to isolate single separable pulses, but it is understandable that other pulses might have been inadvertently included, in particular faint pulses.

It does appear, however, that some pulses are just not good fits to the ESP paradigm.  It is interesting to wonder if all the conjectures that go into the Norris pulse fit are violated, if some of them are violated, or perhaps just one goes awry. Some conjectures may also be more true, or better approximations, than others.

From the sample fit, it is speculated here that the pulse start conjecture is the strongest approximation, the pulse scale conjecture is the next strongest, and the pulse shape conjecture is the weakest, in particular between different pulses.  In other words, it appears that some GRB pulses are better fit with $\xi \ne 1$, but that this shape may be consistent across energy bands for that pulse.  Additionally, it appears that other pulses just cannot be fit with any $\xi$, being, for example, too sharply peaked in the center.

For well fit ESPs, it is hoped that this work has better quantified not only how $A$ and $\tau$ change with energy, but how better known parameters, such as peak flux (proportional to $A$) and fluence (proportional to $A \tau$) change with energy.  When a GRB ESP sample with actual redshifts becomes known, it is hoped that this energy-dependent correlations might one day lead to lower dispersion standard measures bolstering the use of GRBs as cosmological probes.

As with much research, this work might just be the tip of a much larger iceberg. Many questions arise that are not yet answered, and might take considerable effort and greater data to answer.  For example, over what energy range does the ESP paradigm work?  Perhaps GRB data over a wider energy band would indicate clear energy cutoffs at both high and low energies where pulse shapes do some something completely different.

The author wishes to acknowledge useful discussions with Jay P. Norris, Jerry T. Bonnell, Demos Kazanas, John Hakkila, Amir Shahmoradi, Daniel Miller, and Justin Holmes.

\label{lastpage}
 
\end{document}